\documentclass[14pt]{article}
\pdfoutput=1
\usepackage{amsmath,amssymb,amsfonts,amsthm,bm,bbm,cancel,wasysym}
\usepackage{epsfig,graphics,graphicx,epstopdf}
\usepackage{array,booktabs,colortbl,colordvi,multirow}
\usepackage{colordvi,color,xcolor}
\usepackage{hyperref}
\usepackage{rotating}
\usepackage{appendix}

\usepackage{verbatim}
\usepackage{cite}
\usepackage{subfig}
\usepackage{setspace}
\usepackage{url}
\usepackage[percent]{overpic}
\usepackage{slashed}
\usepackage{authblk}
\usepackage{xspace}
\usepackage{fullpage}
\usepackage{hyperref}

\def\ie{{\it i.e.}}
\def\eg{{\it e.g.}}
\def\etc{{\it etc}}

\def\to{\rightarrow}

\newskip\zatskip \zatskip=0pt plus0pt minus0pt
\def\matth{\mathsurround=0pt}
\def\lsim{\mathrel{\mathpalette\atversim<}}

\def\atversim#1#2{\lower0.7ex\vbox{\baselineskip\zatskip\lineskip\zatskip
  \lineskiplimit 0pt\ialign{$\matth#1\hfil##\hfil$\crcr#2\crcr\sim\crcr}}}

\begin{document}

\begin{flushright}
SLAC-PUB-17566\\
\today
\end{flushright}
\vspace*{5mm}

\renewcommand{\thefootnote}{\fnsymbol{footnote}}
\setcounter{footnote}{1}

\begin{center}

{\Large {\bf Building Kinetic Mixing From Scalar Portal Matter}}\\

\vspace*{0.75cm}

{\bf Thomas D. Rueter and Thomas G. Rizzo}~\footnote{tdr38@stanford.edu, rizzo@slac.stanford.edu}

\vspace{0.5cm}

{SLAC National Accelerator Laboratory}\\ 
{2575 Sand Hill Rd., Menlo Park, CA, 94025 USA}

\end{center}
\vspace{.5cm}

\begin{abstract}
 
\noindent
The nature of dark matter (DM) and how it might interact with the particles of the Standard Model (SM) is an ever-growing mystery. It is possible that the existence of new `dark sector' forces, yet undiscovered, are the key to solving this fundamental problem. In this paper, we construct a model in which a dark photon mediates interactions with the SM via kinetic mixing. Unlike traditional models, in which the dark photon, which couples to a dark charge, $Q_D$, mixes with the hypercharge boson, our model effectively mixes the dark photon directly with the photon after electroweak symmetry is broken, but remains unmixed until the symmetry breaks. The kinetic mixing is generated at one loop by fields which satisfy $\sum Q_D Q_{em} = 0$, a condition which guarantees a finite result at one loop. In the literature, this has been traditionally obtained via heavy fermions, which may lie out of the reach of current accelerators. In this model, by contrast, this process is mediated by scalar `portal matter' fields, which are charged under the SU$(2)_L \times $U$(1)_Y$ of the standard model as well as the dark gauge group U$(1)_D$ and acquire GeV-scale vevs which give mass to the dark Higgs and dark photon. The additional scalar fields are relatively light, at or below the weak scale, yet may remain undetected by current experiments since their couplings to SM fermions come only through percent level mixing with the SM Higgs. At colliders, these models are typified by relatively low MET due the BSM states decaying into MET and SM bosons, with MET which is balanced by the decay of the associated production object. Nevertheless, the higher statistics of HL-LHC may be able to probe the entirety of the model space.
\end{abstract}

\renewcommand{\thefootnote}{\arabic{footnote}}
\setcounter{footnote}{0}
\thispagestyle{empty}
\vfill
\newpage
\setcounter{page}{1}

\newpage

\section{Introduction}

Dark matter (DM) makes up a large portion of the matter budget of the universe, as inferred from CMB measurements \cite{aghanim2018planck}, yet it has eluded all attempts at detection to date. As the traditional WIMP parameter space has been probed more extensively \cite{akerib2017results,agnes2018darkside,xenon1t2019}, null results have prompted model building efforts to turn to scenarios in which the DM may be lighter than the traditional WIMP candidate ($\ie$ $m_{DM}\lesssim \mathcal{O}$(GeV)) and/or may be part of a more complex dark sector. Models of dark sectors vary in complexity from the relatively simple addition of a dark gauge group U$(1)_D$ to Twin Higgs models which contain "dark" copies of the entire Standard Model  (SM) \cite{chacko2006natural}, but experiments are only sensitive to the ``portal" interactions which connect the SM to the DM sector. As a result, much of the model building and experimental focus has been on relatively simple models which may constrain the parameter space of more complex theories. In particular, models featuring a dark photon which kinetically mixes (KM) with the U$(1)_Y$ gauge boson have been the object of intense study due to experimental accessibility and the relatively small set of parameters which can robustly produce the observed relic density of DM \cite{vectorportal,vectorportal2}. 

The strength of KM between U$(1)$ gauge bosons depends on the number of loops in the diagrams responsible for the mixing.  A naive estimate of the strength of the KM, 
$\epsilon$, arising from a Dirac fermion with charges ($q_1$, $q_2$) under U$(1)_1\times$ U$(1)_2$ at the one loop level is \cite{holdomKM}
\begin{equation}
\epsilon = \frac{g_1 g_2}{12\pi^2}q_1 q_2 \textrm{ln} \left(\frac{m^2}{\mu^2} \right).
\end{equation}
Depending on the field content of the theory and the mass scale separation in the logarithm, this gives $\epsilon \sim (10^{-2}$-$10^{-1})\times g_1g_2$ in the absence of any more precise cancellations. Experimental searches for dark photons with masses $m_V \gtrsim \mathcal{O}$(10 MeV) constrain $\epsilon \lesssim 10^{-3}$, thus in order to further suppress $\epsilon$ it becomes necessary to induce $\epsilon$ at a higher loop order \cite{gherghettatinyKM} or to arrange charges of the field content such that the sum over the field content $\sum_i q_{1,i}q_{2,i} = 0$, so that $\epsilon \sim \textrm{ln}(m_i^2/m_j^2)$, which may be small for somewhat degenerate masses. The benefit of the latter approach is that while the portal coupling $\epsilon$ may become suppressed beyond the reach of present experiments, the SM charges of the matter in the loop provide a complementary set of search techniques for these theories at colliders \cite{Rizzo:2018vlb,e6portal,kim2020maverick,dolle2010dilepton,gustafsson2012status, gabrielli2014exponential, gabrielli2017radiative, acuna2020phenomenological}.

In this paper we will study a model in which the SM gauge group is extended by an additional U$(1)_D$, with gauge coupling $g_D$, which is broken by a pair of Higgs vevs,  
$v_i$ (with $v_1^2+v_2^2=v_D^2$), at the GeV scale. In order to produce finite kinetic mixing between U$(1)_D$ and U$(1)_Y$, we introduce as ``portal matter" two Higgs doublets $\eta_{1,2}$ which have the same SM quantum numbers as the SM Higgs doublet, but are oppositely charged under the dark gauge group, $\ie$ $Q_D(\eta_1) = - Q_D(\eta_2) = 1$ while the usual SM Higgs has $Q_D=0$. In this setup, $\epsilon$ arises from loops of $\eta_{1,2}$, and the condition $\sum_i Y_i Q_{D,i} = 0$ guarantees a finite value for $\epsilon$ at one loop. Intriguingly, since $v_{SM}$ is the largest vev in the model the additional Higgs fields have masses at or below the weak scale, so that they are well within reach of present colliders. Section \ref{sec:model} of the paper discusses the model setup and the particle content of the theory. Section 3 studies constraints on the parameter space arising from both theory and experiment. Section 4 focuses on collider-oriented signatures of the portal matter, and Section 5 summarizes the results and conclusions.

\section{Model Setup} \label{sec:model}

The goal of this study is to introduce additional scalar fields which are charged under both U$(1)_Y$ and a new gauge group U$(1)_D$ which may mediate dark matter interactions with the Standard Model. In particular, if the additional field content, referred to hereafter as portal matter (PM), has charge assignments which satisfy the relation $\sum_i Y_i Q_{D,i}=0$, then the 
kinetic mixing between the U$(1)_Y$ and U$(1)_D$ fields at one loop will be finite and calculable. In order to break U$(1)_D$, and thus produce a mass for the dark photon, at least one of the fields charged under the dark gauge group must acquire a vacuum expectation value, and in a minimal model this may be one of the PM fields. If the PM fields transform as singlets under SU$(2)_L$, then a vev for any of the PM fields would break U$(1)_{EM}$ as well as U$(1)_D$, so we consider the case of PM fields which are SU$(2)_L$ doublets. In order to maintain $\sum_i Y_i Q_{D,i}=0$, we must add a minimum of two dark doublets to the theory, an extension of the SM which has been previously studied in, \eg, the context of the electroweak phase transition \cite{ahriche2015effects}.

Thus motivated by this minimalist approach, the SM Higgs sector is extended to include two additional Higgs doublets, $\eta_{1,2}$, with ``dark" charges which will play the role of PM. These dark doublets have the same SM quantum numbers as the SM Higgs, but are oppositely charged under U$(1)_D$ so that under SU$(2)_L \times $U$(1)_Y \times $U$(1)_D$ they transform as $\eta_1 \sim (2,\frac{1}{2},1)$ and $\eta_2\sim (2,\frac{1}{2},-1)$ thus maintaining the condition $\sum_i Y_i Q_{D,i}=0$. Denoting the SM Higgs by $\Phi$, the Lagrangian for the scalar sector of the theory becomes

\begin{equation}
\mathcal{L} = (D^\mu \Phi)^\dagger D_\mu \Phi + (D^\mu \eta_1)^\dagger D_\mu \eta_1 +(D^\mu \eta_2)^\dagger D_\mu \eta_2 - U,
\end{equation}
where the covariant derivative can be written as $D_\mu = \partial_\mu - i g (\sigma^j/2) W^j_\mu - i g' Y B_\mu -i g_D Q_D V_\mu$, with $\sigma^{1,2,3}$ being the Pauli matrices. The addition of the dark doublets $\eta_{1,2}$ introduces new terms to the Higgs potential, $U$, which is given by

\begin{equation} \label{higgspot}
\begin{aligned}
U = & \mu^2 \Phi^\dagger \Phi + \mu_1^2 \eta_1^\dagger \eta_1 + \mu_2^2 \eta_2^\dagger \eta_2 +\lambda_1 ( \Phi^\dagger \Phi)^2 +\lambda_{21} ( \eta_1^\dagger \eta_1 )^2  +\lambda_{22} ( \eta_2^\dagger \eta_2 )^2 + \lambda_{31}  \Phi^\dagger \Phi \eta_1^\dagger \eta_1 + \lambda_{32} \Phi^\dagger \Phi \eta_2^\dagger \eta_2 \\ & + \lambda_{41} \Phi^\dagger \eta_1 \eta_1^\dagger \Phi + \lambda_{42} \Phi^\dagger \eta_2 \eta_2^\dagger \Phi + \lambda_5 \Phi^\dagger \eta_1 \Phi^\dagger \eta_2 + \lambda_5^* \eta_1^\dagger \Phi \eta_2^\dagger \Phi + \lambda_6 \eta_1^\dagger \eta_1 \eta_2^\dagger \eta_2 + \lambda_7 \eta_1^\dagger \eta_2 \eta_2^\dagger \eta_1\,.
\end{aligned}
\end{equation}
Using a relative phase between $\Phi$ and $\eta_{1,2}$ we can absorb the phase of $\lambda_5$, so that all of the Higgs potential parameters are real and contribute no additional explicit CP violation in the model. 

From the above it is clear that if neither of $\eta_{1,2}$ acquire a vacuum expectation value (vev), then U$(1)_D$ will remain unbroken, and the dark photon, $V_\mu$,  will remain massless. One might imagine giving mass to the dark photon by the addition of a SM singlet which carries only U$(1)_D$; however if this is the only non-SM vev in the model then one can show that the lightest of the $\eta_{1,2}$ components would be stable. Even if this lightest $\eta_{1,2}$ state is electrically neutral, and thus a DM candidate, it will couple directly to the $Z$ due to the SU(2)$_L\times$U(1)$_Y$ charges of $\eta_{1,2}$. Via this $Z$ coupling this weak-scale DM candidate would have already been detected in direct detection 
experiments \cite{akerib2017results,agnes2018darkside,xenon1t2019}, however, so we must consider an alternative scenario. 

If instead only one of $\eta_{1,2}$ develop a vev, then tadpoles are induced by the $\lambda_5$ term, and the potential is only minimized if $\lambda_5=0$. However, in the absence of $\lambda_5$ the potential develops a Peccei-Quinn symmetry \cite{peccei1977CP,peccei1977constraints}, analogous to the Peccei-Quinn symmetry of the Two Higgs Doublet Model in the absence of soft $Z_2$ breaking terms \cite{ginzburg2005symmetries}. As a result, when $\lambda_5=0$ there is an additional massless neutral pseudoscalar mode in the spectrum beyond the two Goldstone modes which are eaten by the $Z$ and $V$ that one expects from the symmetry breaking pattern, ruling out this scenario. 

Due to the constraints outlined above, we are then forced to consider the case where the neutral components of both $\eta_{1,2}$ develop vevs $v_{1,2}$. We take the vevs to be real, deferring the study of spontaneous CP violation within this model to future work. In this scenario, no stable particles remain, and U$(1)_D$ is broken as desired. Taking $v_{1,2} \sim \mathcal{O}$(GeV) then gives the dark photon a mass near or below the GeV scale. The SM Higgs acquires its usual vev, $v$, and gives mass to the SM fermions, while their non-zero  U$(1)_D$ charges forbid the PM doublets from coupling to the SM fermions, thus avoiding possible tree-level flavor-changing neutral currents in the Higgs sector.

We define the real and imaginary parts of the complex fields as
\begin{equation}
H = \begin{pmatrix} H^+ \\ \frac{h+v+ia}{\sqrt{2}} \end{pmatrix} ~~~
\eta_1 = \begin{pmatrix} \eta_1^+ \\ \frac{\chi_1+v_1+i \xi_1}{\sqrt{2}} \end{pmatrix} ~~~
\eta_2 = \begin{pmatrix} \eta_2^+ \\ \frac{\chi_2+v_2+i \xi_2}{\sqrt{2}} \end{pmatrix}.
\end{equation}
In the absence of U$(1)_D$ breaking (\ie, $v_1=v_2=0$), the dark charge will be a good quantum number, and the states $\eta_1^0$ and $\eta_2^{0*}$ will mix, with this mixing mediated by the $\lambda_5$ term of equation \ref{higgspot}. In the basis of the real fields, this term will mix $\chi_1$ with $\chi_2$ and $\xi_1$ with $-\xi_2$, up to correction terms of order $v_{1,2}/v \simeq 10^{-2}$.  In the absence of CP violation, there are three neutral CP-even scalars, one neutral CP-odd scalar, and 2 charged scalars remaining in the physical spectrum after spontaneous symmetry breaking.

\subsection{CP-Odd Sector}

In the absence of CP-violation, the CP-odd sector contains a single massive state and two Goldstone bosons which are eaten by the SM $Z$ and the dark photon, $V$. The mass matrix is given in the $a$, $\xi_1$, $\xi_2$ basis by

\begin{equation}
M_{\textrm{CPO}}^2 = v^2 \begin{pmatrix} -2 \lambda_5  x_1 x_2 & \lambda_5 x_2 & \lambda_5 x_1 \\ \lambda_5 x_2 & -\frac{\lambda_5 x_2}{2 x_1} & -\frac{\lambda_5}{2} \\ \lambda_5 x_1 & -\frac{\lambda_5}{2} & -\frac{\lambda_5 x_1}{2 x_2} \end{pmatrix},
\end{equation}
where $x_{i} \equiv v_i/v$. We denote the physical, massive CP-odd state by $A$, and find
\begin{equation}
m_A^2= - \frac{\lambda_5 v^2}{2} \left[t+\frac{1}{t}+4x_1 x_2 \right],
\end{equation}
where $t \equiv x_1/x_2 \simeq 1$. In order to have a positive mass squared, we require the product $-\lambda_5 v_1 v_2 >0$, and for the sake of concreteness we will assume $v_{1,2} > 0$ and $\lambda_5<0$ in what follows. The physical field $A$ is an admixture of the pseudoscalar $a$ component of the SM Higgs as well as the pseudoscalar components of $\eta_{1,2}^0$, which we denoted by  $\xi_{1,2}$. Specifically, one finds that $A$ is the admixture 
\begin{equation}
A = \frac{-2 x_1 a + \xi_1 + t\xi_2}{\sqrt{t^2 + 1 +4x_1^2 }}.
\end{equation}

The corresponding Goldstone modes are then linear combinations of the two remaining fields which are orthogonal to $A$:

\begin{equation}
\begin{aligned}
G_Z^0 &= c_{\theta_G} ~G_1 - s_{\theta_G} ~G_2, \\
G_V^0 &= s_{\theta_G} ~G_1 + c_{\theta_G} ~G_2; \\
\vspace{0.2cm}
G_1 &= \frac{a+2x_2 \xi_2}{\sqrt{1+4x_2^2}}, \\
G_2 &=  \frac{2x_2a+(t+4x_1 x_2)\xi_1 - \xi_2}{\sqrt{(1+4x_2^2)(1+t^2+4x_1^2)}},
\end{aligned}
\end{equation}
where $s_{\theta_G}= \sin\theta_G$, \etc, with the angle $\theta_G$ given by

\begin{equation}
\textrm{sin}\theta_G = \frac{-\sqrt{x_1^2+x_2^2+4x_1^2 x_2^2}\left[ \frac{g}{c_w} (1+2x_2^2) - 8 g_D^2 x_2^2 + 2\frac{g g_D \epsilon_{ZV}}{c_w} \right]}{\frac{g^2}{c_w^2}(1-4x_2^2)-(\frac{g^2}{c_w^2} + 4g_D^2)(x_1^2+x_2^2)} +\mathcal{O}(\epsilon^2),
\end{equation}
where $\epsilon_{ZV}$, discussed further in Sec. \ref{sec:gaugekm}, parameterizes the effective kinetic mixing between the dark photon, $V$, and the $Z$, and is of order $\epsilon_{ZV} \sim \epsilon \sim x_i^2 \sim 10^{-4}$. $G_{V,Z}^0$ are the Goldstones eaten by the $V$ and $Z$, respectively. Noting that sin$\theta_G \sim x_i$, we see that $G_{Z}^0$ is primarily composed of the $a$, while $G_V^0$ is primarily an admixture of $\xi_{1,2}$ as might be expected. We also see that $t$ controls the relative amount of $\xi_1$ and $\xi_2$ in the dark photon's Goldstone partner, with $t>1$ increasing the $\xi_1$ admixture and $t<1$ increasing the $\xi_2$ admixture. This may have been expected since $t$ is the ratio of the dark vevs, and $t>1$ reflects the case in which U$(1)_D$ breaking and the dark photon mass are dominated by $v_1$ while $t<1$ implies that $v_2$ dominates the dark photon mass and U$(1)_D$ breaking.

\subsection{Charged Sector}

In the charged sector, there are two physical states, denoted $H_{1,2}^\pm$, and a Goldstone mode, $G^\pm$, which is eaten by the $W^\pm$. These are admixtures of the gauge eigenstates $H^\pm$, the charged component of the SM Higgs, and $\eta_{1,2}^\pm$. The mass matrix in the $H^\pm$, $\eta_1^\pm$, $\eta_2^\pm$ basis is

\begin{equation}
M_{\textrm{ch}}^2 = \frac{v^2}{2}\begin{pmatrix} -\lambda_{41} x_1^2 - \lambda_{42} x_2^2 - 2 \lambda_5 x_1 x_2 & \lambda_{41} x_1 + \lambda_5 x_2 & \lambda_{42} x_2 + \lambda_5 x_1 \\  \lambda_{41} x_1 + \lambda_5 x_2 & -\frac{\lambda_{41} x_1 + \lambda_5 x_2 + \lambda_7 x_1 x_2^2}{x_1} & \lambda_7 x_1 x_2 \\ \lambda_{42} x_2 + \lambda_5 x_1 & \lambda_7 x_1 x_2 & -\frac{\lambda_{42} x_2 + \lambda_5 x_1 + \lambda_7 x_1^2 x_2}{x_2}   \end{pmatrix}.
\end{equation}
The mass eigenstates can be expressed as

\begin{equation}
\begin{aligned}
G^\pm &= \frac{H^\pm + x_1 \eta_1^\pm +x_2 \eta_2^\pm}{\sqrt{1+x_1^2+x_2^2}}, \\
H_1^\pm &= \frac{c_\alpha \sqrt{1+x_2^2}}{\sqrt{1+x_1^2+x_2^2}}\eta_1^\pm + \left[\frac{x_2 s_\alpha}{\sqrt{1+x_2^2}}-\frac{x_1 c_\alpha}{\sqrt{(1+x_2^2)(1+x_1^2+x_2^2)}} \right]H^\pm  - \left[\frac{s_\alpha}{\sqrt{1+x_2^2}} -\frac{x_1 x_2 c_\alpha}{\sqrt{(1+x_2^2)(1+x_1^2+x_2^2)}} \right]\eta_2^\pm, \\
H_2^\pm &= \left[ \frac{c_\alpha}{\sqrt{1+x_2^2}} - \frac{x_1x_2 s_\alpha}{\sqrt{(1+x_2^2)(1+x_1^2+x_2^2)}} \right]\eta_2^\pm  - \left[ \frac{x_2 c_\alpha}{\sqrt{1+x_2^2}}+\frac{x_1 s_\alpha}{\sqrt{(1+x_2^2)(1+x_1^2+x_2^2)}} \right] H^\pm +  \frac{s_\alpha \sqrt{1+x_2^2}}{\sqrt{1+x_1^2+x_2^2}} \eta_1^\pm,
\end{aligned}
\end{equation}
where the angle $\alpha$ is given by 
\begin{equation}
\begin{aligned}
\textrm{tan}(2\alpha) =& \frac{-2x_1x_2^2\sqrt{1+x_1^2+x_2^2}[x_2 \lambda_5 +x_1(\lambda_{41}-\lambda_7)]}{x_2(1+x_1^2+x_2^2)[x_2\lambda_5+x_1(\lambda_{41}+x_2^2\lambda_7)]-x_1[x_2(1+x_2)^2\lambda_{42}+x_1\lambda_5(1+2x_2^2+2x_2^4)+x_1^2x_2(x_2^2\lambda_{41}+\lambda_7)]} \\
\simeq& \frac{2x_2^2\lambda_5+2x_1x_2(\lambda_{41}-\lambda_7)}{\lambda_{42}-\lambda_{41}+\lambda_5(t-\frac{1}{t})} + \mathcal{O}(x_i^4).
\end{aligned}
\end{equation}
The $G^\pm$ is primarily composed of the charged component of the SM Higgs, with $\mathcal{O}(x_i)$ admixtures of $\eta_{1,2}^\pm$, while $H_{1,2}^\pm$ are primarily composed of $\eta_{1,2}^\pm$, respectively, with an $\mathcal{O}(x_i)$ admixtures of the $H^\pm$ and $\mathcal{O}(x_i^2)$ admixtures of $\eta_{2,1}^\pm$. Keeping only terms to order $x_i^2$, the masses for $H_{1,2}^\pm$ are given by

\begin{equation}
\begin{aligned}
m_1^2 &= \left[-\frac{\lambda_{41}}{2}(1+x_1^2) - \frac{\lambda_5}{2 t}(1+x_1^2) - \frac{\lambda_7}{2}x_2^2 \right]v^2 \\
m_2^2 &= \left[-\frac{\lambda_{42}}{2}(1+x_2^2) - \frac{\lambda_5 t}{2}(1+x_2^2) - \frac{\lambda_7}{2}x_1^2 \right]v^2.
\end{aligned}
\end{equation}
Inspecting these expressions, we see that $m_{1,2}^2 \gtrsim m_h^2$ requires $\lambda_{41},\lambda_{42} < 0$ in addition to the requirement $\lambda_5<0$ arising from $m_A^2 > 0$.

\subsection{CP-Even Sector}

The most complicated sector is that of the CP-even neutral fields, with three physical states which are admixtures of $h$, the would-be SM Higgs boson, and the real parts of $\eta_{1,2}^0$, denoted by $\chi_{1,2}$. The mass matrix in the $h, \chi_1,\chi_2$ basis is then 

\begin{equation}
M_{\textrm{CPE}}^2 =v^2\begin{pmatrix}
2 \lambda_1 & x_1(\lambda_{31}+\lambda_{41}) + x_2 \lambda_5 & x_2(\lambda_{32}+\lambda_{42}) + x_1 \lambda_5 \\ 
x_1(\lambda_{31}+\lambda_{41}) + x_2 \lambda_5 & 2\lambda_{21}x_1^2-\frac{\lambda_5}{2t} & \frac{\lambda_5}{2} + x_1 x_2(\lambda_6 + \lambda_7) \\ 
x_2(\lambda_{32}+\lambda_{42}) + x_1 \lambda_5 & \frac{\lambda_5}{2} + x_1 x_2(\lambda_6 + \lambda_7) & 2\lambda_{22}x_2^2-\frac{ t\lambda_5}{2}
\end{pmatrix}.
\end{equation}
Note that the mass mixings between $h$ and $\chi_{1,2}$ are $\mathcal{O}(x_i)\sim10^{-2}$, while the mass mixings within the ``dark" sector between $\chi_{1}$ and $\chi_2$ are $\mathcal{O}(1)$. This hierarchical mixing can be leveraged to make a very good approximation of the required diagonalization process analytically. First we diagonalize the lower right $2\times2$ block, with a large mixing angle $\theta$ given by

\begin{equation} \label{theta}
\textrm{tan}(2\theta)=\frac{2(c-(\lambda_6+\lambda_7)x_1 x_2)}{c(\frac{1}{t}-t) + 2\lambda_{21}x_1^2 -2\lambda_{22}x_2^2}.
\end{equation}
where $x_i$ and $t$ are defined as above, and we introduce $c=-\lambda_5/2 >0$. Under the exchange $t \leftrightarrow \frac{1}{t}$, tan(2$\theta$) will change sign, but this exchange is equivalent to interchanging the labels of $\eta_1$ and $\eta_2$, so for concreteness in the remainder of the paper we will consider the case $t\geq 1$ so that tan$(2\theta) < 0$.  At leading order we may drop the $\mathcal{O}(x_i^2)$ terms in tan($2\theta$), and make the identification

\begin{equation} \label{cstheta}
\textrm{cos}(\theta) \approx \frac{t}{\sqrt{1+t^2}} ~~~~~~~~ \textrm{sin}(\theta) \approx \frac{-1}{\sqrt{1+t^2}}.
\end{equation}
Diagonalizing the rest of the matrix and neglecting terms of $\mathcal{O}(x^3)$, we arrive at the admixtures for the physical states, which are given by

\begin{equation}
\begin{aligned}
h_{SM} =& c_{\theta_1} c_{\theta_2} h + (-c_{\theta} c_{\theta_2} s_{\theta_1}-s_{\theta} s_{\theta_2}) \chi_1+(c_{\theta_2} s_{\theta} s_{\theta_1} - c_{\theta} s_{\theta_2})\chi_2 \\
h_d = (c_{\theta_3} s_{\theta_1}-c_{\theta_1} s_{\theta_2} s_{\theta_3}) h+&(c_{\theta} c_{\theta_1} c_{\theta_2} - c_{\theta_2} s_{\theta} s_{\theta_3} + c_{\theta} s_{\theta_1} s_{\theta_2} s_{\theta_3})\chi_1 + (-c_{\theta_1} c_{\theta_3} s_{\theta} - c_{\theta} c_{\theta_2} s_{\theta_3} - s_{\theta} s_{\theta_1} s_{\theta_2} s_{\theta_3})\chi_2\\
H =  (c_{\theta_1} c_{\theta_3} s_{\theta_2} + s_{\theta_1} s_{\theta_3})h + &(c_{\theta_2} c_{\theta_3} s_{\theta} + c_{\theta} c_{\theta_1} s_{\theta_3} - c_{\theta} c_{\theta_3} s_{\theta_1} s_{\theta_2})\chi_1+(c_{\theta} c_{\theta_2} c_{\theta_3}+ c_{\theta_3} s_{\theta} s_{\theta_1} s_{\theta_2} - c_{\theta_1} s_{\theta} s_{\theta_3})\chi_2,
\end{aligned}
\end{equation}
where the various angles are of order $s_\theta(=\sin (\theta)$, \etc), $c_\theta \sim 1$, $s_{\theta_1}, s_{\theta_2} \sim x_i$ and $s_{\theta_3} \sim x_i^2$.  We identify the $\simeq 125$ GeV,  SM-like Higgs boson with the suggestively named $h_{SM}$, and note that only small admixtures of $h$, of order $x_i\sim10^{-2}$, appear in the other neutral CP-even states. In terms of the Higgs potential parameters and the angle $\theta$ of Eq. \ref{theta}, and dropping $\mathcal{O}(x_i^3)$ terms, these three angles are given by

\begin{equation}
\begin{aligned}
\textrm{tan}(2\theta_1) &= \frac{-c_\theta[x_1(\lambda_{31}+\lambda_{41})+x_2\lambda_5] +s_\theta[x_2(\lambda_{32}+\lambda_{42})+x_1\lambda_5)]}{\lambda_1}, \\
\textrm{tan}(2\theta_2) &= \frac{-2 (c_\theta[x_2(\lambda_{32} + \lambda_{42})+x_1\lambda_5] +s_\theta[x_1(\lambda_{31}+\lambda_{41})+x_2\lambda_5])}{2\lambda_1- c(\frac{1}{t}+t)}, \\
\textrm{tan}(2\theta_3) &= \frac{-s_{\theta_1} t_{2\theta_2}(2\lambda_1-c(\frac{1}{t}+t))}{c(\frac{1}{t}+t)}.
\end{aligned}
\end{equation}
The masses of the physical states can be expressed in terms of the Higgs potential parameters, dropping terms of $\mathcal{O}(x_i^4)$, as 

\begin{equation}
\begin{aligned}
\frac{m_{h_{SM}}^2}{v^2} &= 2\lambda_1(1-2 s_{\theta_1}^2-2 s_{\theta_2}^2) +\lambda_1 s_{2\theta_1} t_{2\theta_1}+ s_{\theta_2}^2 M_3+s_{\theta_2} t_{2\theta_2}(2\lambda_1-c \left(\frac{1}{t}+t \right)),\\
\frac{m_{h_d}^2}{v^2} &= M_2(1-2s_{\theta_1}^2)+2\lambda_1 (s_{\theta_1}^2 - s_{\theta_1} t_{2\theta_1}), \\
\frac{m_{H}^2}{v^2} &=M_3(1-2s_{\theta_2}^2)+2 \lambda_1 s_{\theta_2}^2 - s_{\theta_2} t_{2\theta_2}(2\lambda_1-c\left(\frac{1}{t}+t \right)),
\end{aligned}
\end{equation}
where we have introduced the abbreviations $M_2\sim\mathcal{O}(x_i^2)$ and $M_3\sim\mathcal{O}(1)$:

\begin{equation}
\begin{aligned}
M_2 &= c_\theta^2 \left(\frac{c}{t}  +2x_1^2 \lambda_{21} \right) - s_{2\theta} (-c+x_1 x_2 (\lambda_6 + \lambda_7)) + s_\theta^2 (ct+2x_2^2 \lambda_{22}), \\
M_3 &=  c_\theta^2 (ct+2x_2^2 \lambda_{22}) + s_{2\theta} (-c+x_1 x_2 (\lambda_6 + \lambda_7)) + s_\theta^2\left(\frac{c}{t}  +2x_1^2 \lambda_{21} \right) \\
 &= c\left(\frac{1}{t}+t \right) + \mathcal{O}(x_i^2).
\end{aligned}
\end{equation}
From this we observe that there is a light state, $h_d$, with a mass near the GeV scale, and a heavy state $H$ with a mass very close to the mass of the pseudoscalar $A$. Neglecting all $\mathcal{O}(x_i^2)$ terms, we see that $m_H^2 = m_A^2 = v^2 c(\frac{1}{t}+t)$, so that this $H-A$ degeneracy is broken only by the small U$(1)_D$ breaking terms.

\subsection{Gauge Bosons and Kinetic Mixing} \label{sec:gaugekm}

\begin{figure}
\vspace*{0.5cm}
\centerline{\hspace{-1cm}\includegraphics{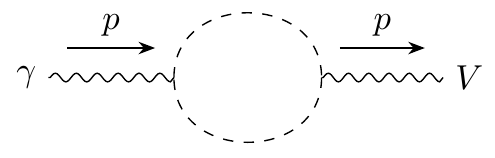} \hspace{1cm}
\includegraphics{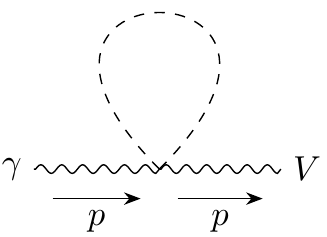}}
\caption{The two, 1-loop diagrams which contribute to $\epsilon$ and $\epsilon_{ZV}$. For $\epsilon$, only the charged Higgses $H_{1,2}^\pm$ run in the loops. For $\epsilon_{ZV}$ we replace $\gamma \rightarrow Z$ and we have the neutral BSM Higgs fields $H$, $A$, $h_d$, and the $V$ Goldstone boson, $G_V^0$, run in the loops in addition to the $H_{1,2}^\pm$.}
\label{kmgraphs}
\end{figure}

Kinetic mixing in this model is somewhat distinct from the usual cases examined in the literature\cite{fabbrichesi2020dark}. In the typical case, kinetic mixing is induced while the SM gauge group remains unbroken, and the dark photon $V$ mixes directly with the hypercharge boson $B$, usually via loops of vector-like fermion PM. The kinetic mixing is then removed by a non-unitary transformation which rescales the dark photon field and couples it to the hypercharge current, and the couplings to mass eigenstates after electroweak symmetry breaking (EWSB) are determined by a standard mass diagonalization procedure. This process can generate finite $\epsilon$ at one loop as long as the portal matter satisfies $\sum_i Y_i Q_{D,i}=0$ and have masses which arise independently of EWSB. The model outlined in section \ref{sec:model} departs from this standard picture, however, as the portal matter masses are themselves generated as a result of the symmetry breaking, by the SM Higgs vev $v$ and/or dark vevs $v_{1,2}$. Since the portal matter states are massless prior to symmetry breaking, they will not generate kinetic mixing in the unbroken theory, and $\epsilon \neq 0$ can only be produced in the broken phase of the theory. 
After symmetry breaking, it is most convenient to consider KM between the usually defined SM fields $A_\mu$, $Z_\mu$, with $V_\mu$ rather than the weak eigenstates $B_\mu$, $W_{3\mu}$, and 
$V_\mu$. Mass mixing between the $Z$ and $V$ will be order $x_i^2 \sim \epsilon$, and thus we will neglect these effects in our estimation of $\epsilon$ itself. Similarly we will only consider the $\mathcal{O}(1)$ mixings of the Higgs bosons which will run in the loop graphs, since the $\mathcal{O}(x_i)$ contributions become $\mathcal{O}(\epsilon x_i)$ terms in the Lagrangian, which are negligible. At leading order, it is convenient to work in the mass eigenstate basis $H$, $A$, $h_d$, $G_V^0$, making the field transformations

\begin{equation} \label{fields}
\begin{aligned}
\chi_1 &\rightarrow c_\theta ~h_d +s_\theta ~H, ~~~~~~~~ \xi_1 \rightarrow c_\theta ~G_V-s_\theta ~A, \\
\chi_2 &\rightarrow -s_\theta ~h_d + c_\theta ~H, ~~~~~~ \xi_2 \rightarrow s_\theta ~G_V+c_\theta ~A,
\end{aligned}
\end{equation}
where we use the leading order estimate for $c_\theta$ and $s_\theta$ given by Eq. \ref{cstheta}. We take $\eta_1^\pm \approx H_1^\pm$ and $\eta_2^\pm \approx H_2^\pm$, as the mixing effects in the charged sector are $\mathcal{O}(x_i)$. Since we are interested in the coupling of the dark photon to SM matter, we focus on kinetic mixing between the SM photon and the dark photon, parameterized by $\epsilon$, and the $Z$ and the dark photon, parameterized by $\epsilon_{ZV}$.  The relevant Feynman diagrams for these are shown in Fig. \ref{kmgraphs}. We denote the kinetically mixed fields with hats, and write the kinetically mixed Lagrangian as 

\begin{equation}
\mathcal{L}_{\textrm{KM}} = -\frac{1}{4}\hat {F}_{\mu \nu} \hat{F}^{\mu \nu} -\frac{1}{4}\hat {Z}_{\mu \nu} \hat{Z}^{\mu \nu} -\frac{1}{4}\hat {V}_{\mu \nu} \hat{V}^{\mu \nu} -\frac{\epsilon}{2}\hat {F}_{\mu \nu} \hat{V}^{\mu \nu} -\frac{\epsilon_{ZV}}{2}\hat {Z}_{\mu \nu} \hat{V}^{\mu \nu}.
\end{equation}

We turn first to the calculation of $\epsilon$, the familiar  $A_\mu-V_\mu$ kinetic mixing parameter. Here only the charged $H_{1,2}^\pm$ contribute in the loop graphs, and they satisfy $\sum_{i=H_{1,2}^\pm} Q_i Q_{D,i} = 0$ so that $\epsilon$ is indeed finite at one loop. We find the familiar-looking result
\begin{equation} \label{eq:eps}
\epsilon= \frac{g_D e}{48 \pi^2}\textrm{ln} \left(\frac{m_2^2}{m_1^2} \right).
\end{equation}

\begin{figure}
\centerline{\includegraphics[width=4.0in,angle=0]{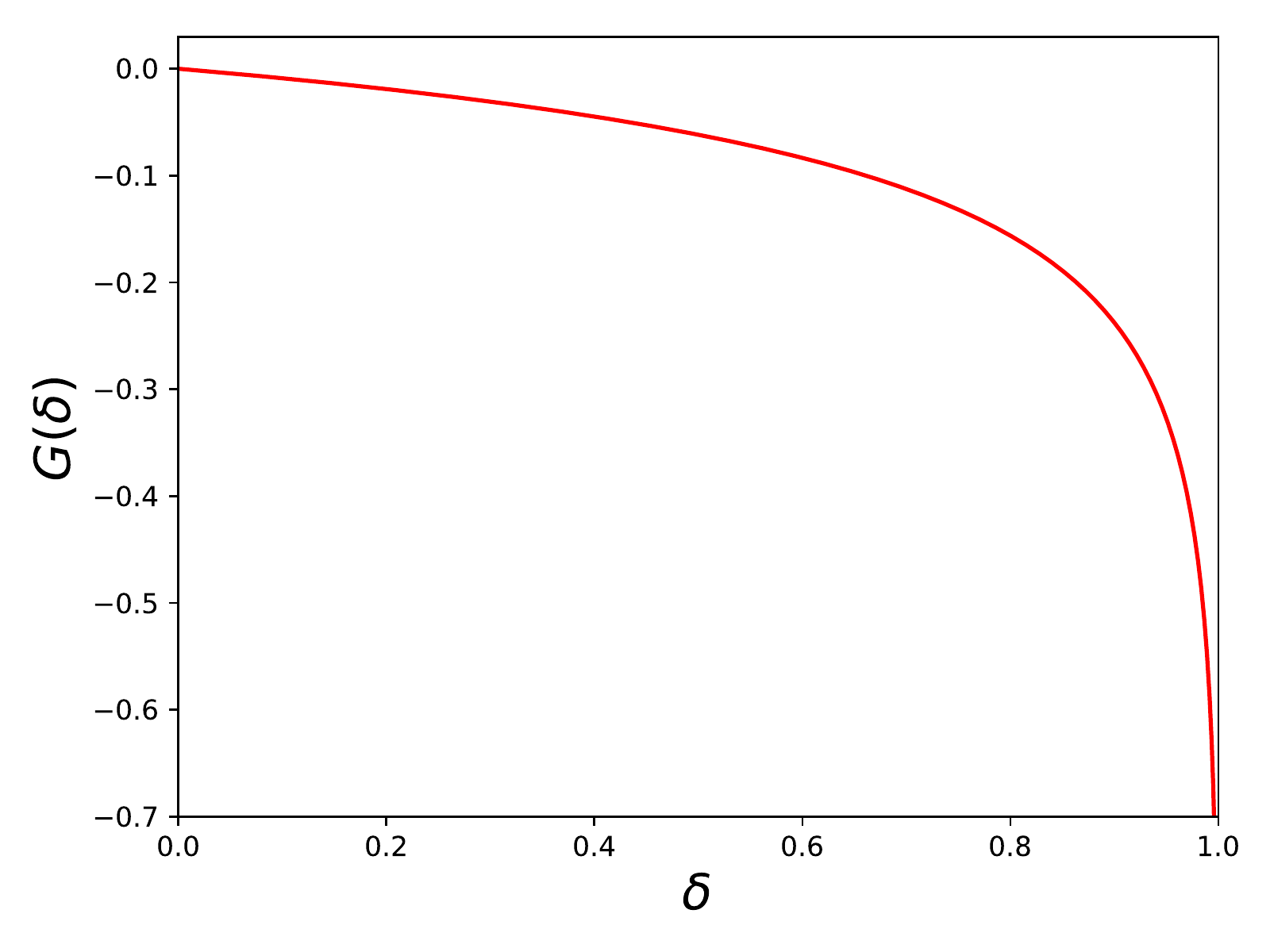}}
\caption{The function $G(\delta)$, as defined in Eq. \ref{gdel}. Typical values of $\delta$ for the models in this paper are $\delta \approx 0.985$, which corresponds to $G(0.985) \approx -0.5$. }
\label{gfig}
\end{figure}

The calculation of $\epsilon_{ZV}$ involves the neutral BSM Higgs bosons as well as the charged states, and we will work in the $\xi=1$ gauge so that the relevant fields in the loop are $H$, $A$, $h_d$, and $G_V^0$. The $H$ and $A$ couple to each other and to $Z_\mu$ and $V_\mu$ via derivative couplings, and to the $Z_\mu$ and $V_\mu$ and themselves via the four point couplings. Since $H$ and $A$ are nearly degenerate, up to a mass splitting $m_H^2 - m_A^2\sim \mathcal{O}(x_i^2 v^2)$, we will assume $m_H \simeq m_A$ so that they contribute to $\epsilon_{ZV}$ as a single neutral complex scalar. The two charged Higgs $H_{1,2}^\pm$ also contribute to $\epsilon_{ZV}$ as complex scalars, just as they did to $\epsilon$. The dark Higgs, $h_d$, and the $V_\mu$ Goldstone, $G_V^0$, couple to $Z_\mu$ and $V_\mu$ similarly to the $H$ and $A$, but they contribute to $\epsilon_{ZV}$ with opposite sign so that the total logarithmic contribution from the set of fields $H$, $A$, $h_d$, and $G_V^0$ is again finite. Since $h_d$ and $G_V^0$ have masses which are set by $\mathcal{O}(x_i v)\sim$ GeV, the mass splitting $m_{h_d}^2-m_V^2 \sim \mathcal{O}(x_i^2v^2)$ is of the same order and produces an additional finite contribution to $\epsilon_{ZV}$. Denoting the fractional mass splitting $\delta = (m_{h_d}^2 - m_V^2)/m_{h_d}^2 \sim \mathcal{O}(1)$, this additional finite contribution to $\epsilon_{ZV}$ is proportional to the function $G(\delta)$ defined below and shown in Fig. \ref{gfig}. For $g_D=e=\sqrt{4 \pi \alpha_{EM}}$, we find the $G(\delta)$ term to be $\sim 25\%$ of the contribution arising from the ln$(m_V^2/m_A^2$) term. We emphasize that $G(\delta)\rightarrow 0$ as $\delta \rightarrow 0$ so that the small fractional mass splitting of the $H$ and $A$, $\delta_{HA} \sim \mathcal{O}(x_i^2)$, may be safely neglected. We thus find $\epsilon_{ZV}$ to be
\begin{equation} \label{eq:epszv}
\epsilon_{ZV} = \frac{g g_D}{48 \pi^2 c_w} \left[ \left( \frac{1}{2} - s_w^2 \right) \textrm{ln}\left( \frac{m_2^2}{m_1^2} \right) + \frac{c_{2\theta}}{2}\left( \textrm{ln} \left( \frac{m_V^2}{m_{A}^2}\right)   - 6 G(\delta) \right)\right],
\end{equation}
where $\delta = (m_{h_d}^2 - m_V^2)/m_{h_d}^2$ and $G(\delta)$, shown in Fig. \ref{gfig}, is given by

\begin{equation} \label{gdel}
G(\delta) = \frac{1}{\delta^3} \left[ \frac{2 \delta (1-\delta)}{3} + \frac{2 \delta^3}{9} + \textrm{ln}(1-\delta)\left( \frac{2}{3} - \delta + \frac{\delta^2}{2} \right) \right].
\end{equation}

To remove both kinetic mixing effects we can make a non-unitary transformation \cite{heeck2011kinetic}

\begin{equation}
\begin{pmatrix} \hat A_\mu \\ \hat Z_\mu \\ \hat V_\mu \end{pmatrix} = \begin{pmatrix} 1 & 0 & -\epsilon/D \\ 0 & 1 & -\epsilon_{ZV}/D \\ 0 & 0 & 1/D \end{pmatrix} \begin{pmatrix} A_\mu \\ Z_\mu \\ V_\mu  \end{pmatrix}, 
\end{equation}
where $D=\sqrt{1-\epsilon^2 - \epsilon_{ZV}^2} \simeq 1$ to leading order. Removing the KM produces mass mixing between the $Z_\mu$ and $V_\mu$ which is $\mathcal{O}(\epsilon_{ZV} v^2)$. This combines with the mass mixing induced at similar order by the dark vevs $v_{1,2}$, which are $\mathcal{O}(x_i^2 v^2)$, to produce a mass matrix in the $Z_\mu$, $V_\mu$ basis given by

\begin{equation}
M_{ZV}^2 = \begin{pmatrix} \frac{g^2 v^2}{4 c_w^2}(1+x_1^2+x_2^2) & -\frac{g_D g v^2}{2 c_w}(x_1^2-x_2^2) - \frac{g^2 \epsilon_{ZV} v^2}{4 c_w^2}(1+x_1^2+x_2^2) \\ -\frac{g_D g v^2}{2 c_w}(x_1^2-x_2^2) - \frac{g^2 \epsilon_{ZV} v^2}{4 c_w^2}(1+x_1^2+x_2^2) & g_D^2 v^2(x_1^2+x_2^2) + \frac{g_D g \epsilon_{ZV} v^2}{c_w} (x_1^2 - x_2^2) + \frac{g^2 \epsilon_{ZV}^2 v^2}{4 c_w^2} \end{pmatrix},
\end{equation}
where we have dropped terms of order $\epsilon^3$. Diagonalizing the mass matrix by a rotation given by

\begin{equation}
\textrm{sin}\theta_{ZV} = \frac{2 g_D c_w}{g}(x_1^2-x_2^2) + \epsilon_{ZV} + \mathcal{O}(\epsilon^2),
\end{equation}
we find the physical masses to be

\begin{equation}
M_Z^2 = v^2\left[ \frac{g^2}{4 c_w^2}(1+x_1^2+x_2^2+\epsilon_{ZV}^2) + g_D^2 (x_1^2-x_2^2)^2 + \mathcal{O}(\epsilon^3)   \right],
\end{equation}
\begin{equation}
M_V^2 = g_D^2v^2 \left[ (x_1^2+x_2^2) - (x_1^2 - x_2^2)^2 + \mathcal{O}(\epsilon^3) \right],
\end{equation}
so the relevant piece of the covariant derivative describing the interactions becomes

\begin{equation}
-ieQA_\mu- i \left[ \frac{g}{c_w}\left(T^3_L - s_w^2 Q \right) - g_D s_{\theta_{ZV}} Q_D \right] Z_\mu -i \left[-\epsilon e Q +2 g_D(x_1^2-x_2^2)\left(T^3_L - s_w^2 Q \right) + g_D Q_D \right] V_\mu,
\end{equation}
where we have now dropped $\mathcal{O}(\epsilon^2)$, \etc,  suppressed terms. We note that the dark photon coupling to the SM is modified from the typical case in the literature due to mass mixing with the $Z$, with a strength determined by the dark sector coupling, $g_D$, (rather than known SM couplings) at leading order. This is similar to the situations observed previously in the literature, where vevs which are charged under both SU(2)$_L\times$U$(1)_Y$ and U(1)$_D$ generate $Z-V$ mass mixing proportional to the ratio of the couplings \cite{e6portal}. Whether the conventional $\epsilon e Q$ term dominates the DP interaction with the SM fields thus depends on the details of the dark sector via the relative size of the product $e \epsilon$ compared to the combination $g_D (x_1^2-x_2^2)$. We note as well that the DP now must also couple to the SM neutrinos due to its mass mixing with the $Z$. Thus $V$ may mediate non-standard neutrino interactions \cite{abdullah2018coherent,lindner2018neutrino}, but we note that these become vanishingly small as $t\to 1$ since the $Z-V$ mass mixing vanishes in this limit.

\section{Constraints on the Model Space} \label{sec:constraint}

\subsection{Higgs Potential} \label{sec:unitarity}

The Higgs potential in Eq. \ref{higgspot} will be minimized by the vevs $v, v_1, v_2$ only if the masses of the various Higgs states in Section \ref{sec:model} are positive. This requirement led to the constraints $\lambda_{41},\lambda_{42},\lambda_5<0$, and the identification of $h_{SM}$ with the SM Higgs sets the additional constraint that $\lambda_1 \simeq 0.129$ so that $m_{h_{SM}} \simeq 125.1$ GeV, up to $\mathcal{O}(x_i^2)$ corrections. Beyond the requirements of positive mass eigenvalues, there are additional theoretical constraints on the couplings $\lambda_i$ arising from unitarity and vacuum stability.

Unitarity constraints on the $\lambda_i$ come from the high energy behavior of 2-to-2 scattering in the Higgs sector, where the dominant contribution to the generic scalar scattering amplitude $S_1 S_2 \rightarrow S_3 S_4$ comes from the quartic terms in the potential. Since at high energies SU$(2)_L\times$U$(1)_Y\times$U$(1)_D$ is unbroken, we may consider the scattering between states of definite hypercharge, dark charge, and isospin. Following the methodology of Ref. \cite{ginzburg2005tree}, we categorize our states as scalar products with $Y= 0, 1,-1$; $\sigma= 0$ (weak isoscalar) or 1 (weak isovector);  and $Q_D = 0, 1,2, -1, -2$. The weak isoscalar states are listed in Table \ref{tab:isoscalars}, while the weak isovector states are listed in Table \ref{tab:isovectors}. The states with $Y=-1$ can be obtained from the $Y=1$ states by conjugation.

\begin{table} 
\begin{center}
\begin{tabular}{ |c |c |c | }
\hline
  & $Y=0$, $\sigma=0$ & $Y=1$, $\sigma=0$: \\
\hline
$Q_D=0$ & $\frac{1}{\sqrt{2}}\Phi^\dagger \Phi,\frac{1}{\sqrt{2}}\eta_1^\dagger \eta_1,\frac{1}{\sqrt{2}}\eta_2^\dagger \eta_2$ & $\frac{1}{\sqrt{2}}\tilde\eta_1 \eta_2$  \\
$Q_D=1$ & $\frac{1}{\sqrt{2}}\Phi^\dagger \eta_1,\frac{1}{\sqrt{2}}\eta_2^\dagger \Phi$ & $\frac{1}{\sqrt{2}}\tilde\Phi \eta_1$  \\
$Q_D=2$ & $\frac{1}{\sqrt{2}}\eta_2^\dagger \eta_1$ & absent  \\
$Q_D=-1$ & $\frac{1}{\sqrt{2}}\eta_1^\dagger \Phi,\frac{1}{\sqrt{2}}\Phi^\dagger \eta_2$ & $\frac{1}{\sqrt{2}} \tilde\Phi \eta_2$  \\
$Q_D=-2$ & $\frac{1}{\sqrt{2}}\eta_1^\dagger \eta_2$ & absent  \\
\hline
\end{tabular}
\end{center}
\caption{The weak isoscalar states which form the gauge eigenstate basis of the potential in Eq. \ref{higgspot} for high energy 2 to 2 scattering.} \label{tab:isoscalars}
\end{table}

\begin{table} 
\begin{center}
\begin{tabular}{|c | c | c|  }
\hline
  & $Y=0$, $\sigma=1$: & $Y=1$, $\sigma=1$: \\
\hline
$Q_D=0$ & $\frac{1}{\sqrt{2}}\Phi^\dagger \tau^i \Phi,\frac{1}{\sqrt{2}}\eta_1^\dagger \tau^i \eta_1,\frac{1}{\sqrt{2}}\eta_2^\dagger \tau^i \eta_2$ &  $\frac{1}{2}\tilde \Phi \tau^i \Phi, \frac{1}{\sqrt{2}}\tilde \eta_1 \tau^i \eta_2=\frac{1}{\sqrt{2}}\tilde \eta_2 \tau^i \eta_1$ \\

$Q_D=1$ & $\frac{1}{\sqrt{2}}\Phi^\dagger \tau^i \eta_1,\frac{1}{\sqrt{2}}\eta_2^\dagger \tau^i \Phi$ &  $ \frac{1}{\sqrt{2}}\tilde \Phi \tau^i \eta_1=\frac{1}{\sqrt{2}}\tilde \eta_1 \tau^i \Phi$  \\
$Q_D=2$ & $\frac{1}{\sqrt{2}}\eta_2^\dagger \tau^i \eta_1$ & $\frac{1}{2}\tilde \eta_1 \tau^i \eta_1$  \\
$Q_D=-1$ & $\frac{1}{\sqrt{2}}\eta_1^\dagger \tau^i \Phi,\frac{1}{\sqrt{2}}\Phi^\dagger \tau^i \eta_2$ & $\frac{1}{\sqrt{2}}\tilde \Phi \tau^i \eta_2=\frac{1}{\sqrt{2}}\tilde \eta_2 \tau^i \Phi$ \\
$Q_D=-2$ & $\frac{1}{\sqrt{2}}\eta_1^\dagger \tau^i \eta_2$ & $\frac{1}{2}\tilde \eta_2 \tau^i \eta_2$  \\
\hline
\end{tabular}
\end{center}
\caption{The weak isovector states which form the gauge eigenstate basis of the potential in Eq. \ref{higgspot} for high energy 2 to 2 scattering.} \label{tab:isovectors}
\end{table}
We note that the U$(1)_D$ charge of the two particle states plays a role analogous to the softly broken $Z_2$ symmetry of the Two Higgs Doublet Model in preventing scattering between two particle states with different $Q_D$ values, though in this instance it is due to a gauge symmetry rather than an imposed discrete symmetry. As a result, it is an instructive check to compare the $Z_2$-odd results of Ref. \cite{ginzburg2005tree} with the $Q_D=\pm1$ results here. Following the notation of Ref. \cite{ginzburg2005tree}, we find the tree-level scattering matrices in the isoscalar channels, $S_{Y,\sigma=0,Q_D}$, to be

\begin{equation}
\begin{aligned}
&16 \pi S_{Y=0, \sigma=0, Q_D=0} = \begin{pmatrix} 6\lambda_1 & 2\lambda_{31}+\lambda_{41} & 2\lambda_{32} + \lambda_{42} \\ 2\lambda_{31} + \lambda_{41} & 6\lambda_{21} & 2\lambda_6+\lambda_7 \\ 2\lambda_{32}+\lambda_{42} & 2\lambda_6+\lambda_7 & 6\lambda_{22}\end{pmatrix}, \\ 
&16 \pi S_{Y=0, \sigma=0, Q_D=1} = \begin{pmatrix} \lambda_{31}+2\lambda_{41} & 3\lambda_5 \\ 3\lambda_5 & \lambda_{32}+2\lambda_{42}      \end{pmatrix},\\
&16 \pi S_{Y=0, \sigma=0, Q_D=2} = \lambda_6+2\lambda_7,\\
&16 \pi S_{Y=1, \sigma=0, Q_D=0} = \lambda_{6}-\lambda_{7},\\
&16 \pi S_{Y=1, \sigma=0, Q_D=1} = \lambda_{31}-\lambda_{41},\\
&16 \pi S_{Y=1, \sigma=0, Q_D=-1} = \lambda_{32} - \lambda_{42},\\
\end{aligned}
\end{equation}
\vspace{0.2cm}

\noindent
where the corresponding matrices for $Y=-1,Q_D=0,\pm1$ and $Y=0,Q_D=-1,-2$, obtained through charge conjugation, will be the same as the $Y=1,Q_D=0,\mp1$ and $Y=0,Q_D=1,2$ cases, respectively, since all parameters in the potential are real. The corresponding tree-level scattering matrices in the isovector channels, $S_{Y,\sigma=1,Q_D}$, are given by

\begin{equation}
\begin{aligned}
&16 \pi S_{Y=0, \sigma=1, Q_D=0} = \begin{pmatrix} 2\lambda_1 & \lambda_{41} & \lambda_{42} \\ \lambda_{41} & 2\lambda_{21} & \lambda_7 \\ \lambda_{42} & \lambda_7 & 2\lambda_{22}    \end{pmatrix},\\
&16 \pi S_{Y=0, \sigma=1, Q_D=1} = \begin{pmatrix} \lambda_{31} & \lambda_5 \\ \lambda_5 & \lambda_{32}  \end{pmatrix},\\
&16 \pi S_{Y=0, \sigma=1, Q_D=2} = \lambda_6\\
&16 \pi S_{Y=1, \sigma=1, Q_D=0} = \begin{pmatrix} 2\lambda_1 & \sqrt{2}\lambda_5 \\ \sqrt{2}\lambda_5 & \lambda_6+\lambda_7 \end{pmatrix},\\
&16 \pi S_{Y=1, \sigma=1, Q_D=1} = \lambda_{31}+\lambda_{41},\\
&16 \pi S_{Y=1, \sigma=1, Q_D=2} = 2\lambda_{21},\\
&16 \pi S_{Y=1, \sigma=1, Q_D=-1} = \lambda_{32}+\lambda_{42},\\
&16 \pi S_{Y=1, \sigma=1, Q_D=-2} = 2\lambda_{22},
\end{aligned}
\end{equation}
\vspace{0.2cm}

\noindent
where again states with $Y=-1$ and/or $Q_D=-1,-2$ may be obtained by charge conjugation, but will be the same as the corresponding matrix above. The unitarity constraint can be written $S_{Y,\sigma,Q_D}<1$, which constrains the absolute values of the eigenvalues of the above matrices, $\Lambda_i$, to satisfy $\left| \Lambda_i \right| < 16\pi$. This produces some constraints on linear combinations of couplings or quadratic functions of couplings for many of the above $(Y,\sigma,Q_D)$ states, but for the 3x3 matrices of $S_{Y=0,\sigma=0,Q_d=0}$ and $S_{Y=0,\sigma=1,Q_d=0}$, these are constraints on cubic equations which translate into complicated constraints on the parameters. We confirm these conditions numerically during our scan of the parameter space. 

The Higgs potential in Eq. \ref{higgspot} must also be bounded from below in order for the minimum characterized by the vevs $v, v_1, v_2 \neq 0$ to be stable. This requirement sets additional constraints on the quartic couplings, and analytic forms of these constraints were found in Ref. \cite{ahriche2015effects} for the case of negligible $\lambda_5$. In the case considered here we cannot neglect $\lambda_5$, as it controls the mass of the pseudoscalar $A$, and $m_A \gtrsim m_h$ requires a sizable $-\lambda_5 \gtrsim 0.3$. 

To ensure the Higgs potential is bounded from below, it is sufficient to demonstrate that the quartic portion of the potential can be written in the form $\lambda_{ab} \phi^2_a \phi^2_b$, where $\phi_{a,b}$ are real fields or gauge orbit variables and $\lambda_{ab}$ is a copositive matrix \cite{kannike2012vacuum}. A symmetric matrix $B$ is copositive if the quadratic form $x^T B x \geq 0$ for all $x \in \mathbb{R}^n_+$, so we simply need to express the potential as $\vec{h}^T \Lambda \vec{h}$ where $\vec{h}$ is a set of non-negative monomials, and demonstrate that $\Lambda$ is copositive. We begin by defining 

\begin{equation}
\begin{aligned}
&\Phi = f \hat{\Phi}, ~~~~~ \eta_i = e_i \hat{\eta_i}, ~~~~~ \hat{\Phi}^\dagger \hat{\Phi} = \hat{\eta_i}^\dagger \hat{\eta_i} = 1, ~~~~ f,e_i > 0, \\
&\hat{\Phi}^\dagger \hat{\eta_i} = \rho_i e^{i \delta_i}, ~~~~ \hat{\eta_1}^\dagger \hat{\eta_2} = \rho' e^{i\phi}, ~~~~ 0\leq \rho_i,\rho' \leq 1.
\end{aligned}
\end{equation}
Using these definitions we may write the quartic terms of Eq. \ref{higgspot} as 

\begin{equation}
\begin{aligned}
V_4 = \lambda_1 f^4 +& \lambda_{21} e_1^4 + \lambda_{22} e_2^4 + \lambda_{31} f^2 e_1^2 + \lambda_{32} f^2 e_2^2 + \lambda_{41} f^2 e_1^2 \rho_1^2 +\lambda_{42} f^2 e_2^2 \rho_2^2 \\ 
&+ 2\lambda_5 f^2 e_1 e_2 \rho_1 \rho_2 \textrm{cos}(\delta_1+\delta_2) + \lambda_6 e_1^2 e_2^2 + \lambda_7 \rho'^2e_1^2 e_2^2.
\end{aligned}
\end{equation}
To ensure vacuum stability,  it is sufficient to minimize the potential with respect to $\rho_i, \rho',$ and $\delta_i$, and show that the resulting matrix representation of the potential $\vec{h}^T\Lambda\vec{h}$, with $\vec{h}^T = (e_1e_2, f^2,e_1^2,e_2^2)$, has copositive $\Lambda$. It is difficult in general to write the minimum of this function for arbitrary values of $\lambda_i$ since the values of $\rho_i, \rho'$, and $\delta_i$ which minimize it are necessarily functions of the parameters $\lambda_i$. However, since we are interested in a particular portion of parameter space where $\lambda_5, \lambda_{41}, \lambda_{42} < 0$, we can minimize with respect to $\rho_i, \rho'$, and $\delta_i$ in a convenient manner. In particular, we may write

\begin{equation}
\lambda_{41} f^2 e_1^2 +\lambda_{42} f^2 e_2^2 + 2 \lambda_5 f^2 e_1 e_2 \leq \lambda_{41} f^2 e_1^2 \rho_1^2 +\lambda_{42} f^2 e_2^2 \rho_2^2 + 2\lambda_5 f^2 e_1 e_2 \rho_1 \rho_2 \textrm{cos}(\delta_1+\delta_2),
\end{equation}
since cos$(\delta_1+\delta_2)=1$ will ensure the $\lambda_5$ term contributes negatively, $\ie ~ 2\lambda_5f^2e_1e_2\rho_1\rho_2 \leq 0$ for $\lambda_5<0$, and $\rho_1 = \rho_2 = 1$ minimizes $\lambda_{41} f^2 e_1^2 \rho_1^2 +\lambda_{42} f^2 e_2^2 \rho_2^2  + 2\lambda_5f^2e_1e_2\rho_1\rho_2$ for $\lambda_{41},\lambda_{42},\lambda_5<0$. Additionally, since $\lambda_7 \rho'^2 e_1^2 e_2^2$ is the only term dependent on $\rho'$, it will be minimized by $\rho=1$ for $\lambda_7<0$ and $\rho'=0$ for $\lambda_7>0$, so that at the minimum the term takes on the value min(0,$\lambda_7) e_1^2 e_2^2$. Thus the minimum of the quartic terms of potential may be written as

\begin{equation} \label{v4min}
V_{4,\textrm{min}} =\frac{1}{2} \begin{pmatrix} e_1 e_2 & f^2 & e_1^2 & e_2^2 \end{pmatrix} \begin{pmatrix} 2c_1\lambda_{67} & 2\lambda_5 & 0 & 0 \\ 2 \lambda_5 & 2 \lambda_1 & \lambda_{31} + \lambda_{41} & \lambda_{32}+\lambda_{42} \\ 0 & \lambda_{31} + \lambda_{41} & 2\lambda_{21} & c_2 \lambda_{67} \\ 0 &   \lambda_{32}+\lambda_{42} & c_2 \lambda_{67} & 2\lambda_{22} \end{pmatrix} \begin{pmatrix} e_1 e_2 \\ f^2 \\ e_1^2 \\ e_2^2 \end{pmatrix},
\end{equation} 
where $\lambda_{67} = \lambda_6 + \textrm{min}(0,\lambda_7)$ and $c_1+c_2=1$. Note that the relationship $c_1+c_2=1$ defines an affine subspace $\Lambda(c_i)$ of the general 4x4 matrix space, and that $V_{4,\textrm{min}}$ is invariant under affine transformations within this subspace. Therefore if any point of the subspace $\Lambda(c_i)$ is copositive, the potential will be bounded from below. There exist copositivity criteria for 4x4 matrices in the literature \cite{andersson1995criteria,ping1993criteria}, but for the present study we satisfy ourselves by confirming that the affine subspace $\Lambda(c_i)$ contains a positive-definite matrix for $c_1 = 1$. Since positive-definite matrices are a subset of copositive matrices \cite{cottle1970copositive}, this condition is sufficient but not necessary to confirm that $V_4$ is bounded below. To check that $\Lambda(c_1=1)$ is positive-definite, we employ Sylvester's Criterion \cite{gilbert1991positive}, which is both necessary and sufficient to show that a Hermitian matrix, such as that of Eq. \ref{v4min}, is positive-definite. 

\subsection{Constraints from Invisible Widths}

Since the dark Higgs and dark photon (eventually) dominantly decay to DM, the presence of these new light states, $V$ and $h_d$, which couple to the SM will introduce new invisible decay channels for both the SM Higgs, $h_{SM}$, and the $Z$. Neglecting $\mathcal{O}(x_i)$ terms, we can take $h \rightarrow h_{SM}$ and use the field identifications of Eq. \ref{fields}, where we then use the leading order estimate for $c_\theta$ and $s_\theta$ given by Eq. \ref{cstheta}. These substitutions in the covariant derivatives of $\eta_{1,2}$ yield a $ZVh_d$ coupling which mediates $Z \rightarrow V h_d$ decays (which are assumed to result in an invisible final state), with partial width
\begin{equation}
\Gamma(Z\rightarrow V h_d) = \frac{g^2 M_z}{96 \pi c_w^2}\frac{c_{2\theta}^2}{2} = \Gamma(Z\rightarrow \nu \bar\nu) \frac{c_{2\theta}^2}{2},
\end{equation}
where we have treated $h_d$ and $V$ as essentially massless, and $\Gamma(Z\rightarrow \nu \bar\nu)$ is the partial width for a single species of neutrino. Writing $c_{2\theta} = (t^2-1)/(t^2+1)$, this invisible width sets a constraint on the allowed values of $t$. Requiring $\Gamma(Z\rightarrow V h_d) \leq 0.0146~ \Gamma(Z\rightarrow \nu \bar\nu)$, a value consistent with a 95\% CL limit on the deviation from the central value of $N_\nu$ as measured from the $Z$ invisible width \cite{pdg2020}, leads us to the condition 
\begin{equation} \label{tcon}
0.8415 \lsim t \lsim 1.1884.
\end{equation}
Since we have chosen to work with $t \geq 1$, this constraint is actually realized as $1 \leq t \lsim 1.1884$. 

Making the replacements of Eq. \ref{fields} into the Higgs potential and using the leading order values for $c_\theta, s_\theta$ in Eq. \ref{cstheta}, we see that $h_{SM}$ couples to the BSM states as 

\begin{equation}
\begin{aligned}
\mathcal{L} \supset -\frac{h_{SM} v}{2} &\bigg\{ (h_d^2 + G_V^2) \left[\frac{t^2}{1+t^2}(\lambda_{31} + \lambda_{41}) + \frac{1}{1+t^2}(\lambda_{32}+\lambda_{42}) + \frac{2t}{1+t^2}\lambda_5\right]  \\
&+(A G_V - H h_d) \left[ \frac{2t}{1+t^2}(\lambda_{31}+\lambda_{41}-\lambda_{32}-\lambda_{42}) - 2\frac{t^2-1}{1+t^2} \lambda_5 \right] \\
&+(H^2 + A^2) \left[ \frac{1}{1+t^2}(\lambda_{31} + \lambda_{41}) + \frac{t^2}{1+t^2}(\lambda_{32}+\lambda_{42}) -\frac{2t}{1+t^2}\lambda_5  \right] \bigg\}.
\end{aligned} 
\end{equation}
These couplings thus mediate new invisible decays $h_{SM}\rightarrow h_d h_d$ and $h_{SM} \rightarrow VV$. By the Goldstone Boson Equivalence Theorem \cite{GBET} we may take $\Gamma(h_{SM}\rightarrow VV) \simeq \Gamma(h_{SM} \rightarrow G_V G_V) = \Gamma(h_{SM} \rightarrow h_d h_d)$ at leading order. In the limit that $m_{h_d}^2/m_h^2 \to 0$, 
we obtain

\begin{equation}
\Gamma(h_{SM} \rightarrow h_d h_d) = \frac{\tilde \lambda_h^2 v^2}{32 \pi m_{h_{SM}}}.
\end{equation}
where we define $\tilde \lambda_h = \frac{t^2}{1+t^2}(\lambda_{31} + \lambda_{41}) + \frac{1}{1+t^2}(\lambda_{32}+\lambda_{42}) + \frac{2t}{1+t^2}\lambda_5$. Searches for invisible Higgs decays at the LHC have recently set a bound on the branching fraction $\mathcal{B}(h_{SM} \rightarrow \textrm{inv.}) < 0.11$ \cite{ATLAShinv},  which translates into a corresponding constraint on the coupling 
\begin{equation} \label{lamtilcon}
|\tilde \lambda_h| < 6.8\times10^{-3}. 
\end{equation}
Note that this constraint forces tan($2\theta_1$) to be small, since at leading order in $x_i$ we can write tan($2\theta_1$) = $-\sqrt{x_1^2+x_2^2}\tilde \lambda_h /\lambda_1 \sim \mathcal{O}(x_i^2)$.

\section{Model Signals} \label{sec:signal}

\subsection{Parameter Scan} \label{sec:scan}

In order to probe the parameter space of the model, we performed a linear flat scan over the 10 $\lambda_i$ parameters of the Higgs potential and $t=x_1/x_2$, setting $x_1$ by taking $v_1 = 1$ GeV. As outlined above, we require $\lambda_5, \lambda_{41}, \lambda_{42} < 0$ to ensure positive masses for the  $A$ and $H_{1,2}^\pm$, respectively. The Higgs potential parameters are required to satisfy $\left| \lambda_i \right| < 5$, and we also require that $1 \leq t \leq 1.1884$ due to the constraint on the invisible width of the $Z$. For each point in parameter space, the unitarity constraints on the Higgs potential of Sec. \ref{sec:unitarity} are verified, as well as the coupling constraint from the invisible width of the SM Higgs of Eq. \ref{lamtilcon}. To increase the number of points which pass the scan, we require $\left| \lambda_{3i}+\lambda_{4i} \right| \leq 2$, since the constraint of Eq. \ref{lamtilcon} relies on these quantities. This increases the efficiency of points passing the scan requirements by a factor of $\sim 8$. Finally, since we expect that light neutral and charged states would have been seen in previous collider searches, we will also require $m_A > 150$ GeV and $m_{1,2} >200$ GeV. A scan of $5\times10^8$ randomly chosen points in parameter space yielded 6884 points which simultaneously satisfied these multiple requirements.

\begin{figure}
\centerline{\includegraphics[width=4.5in,angle=0]{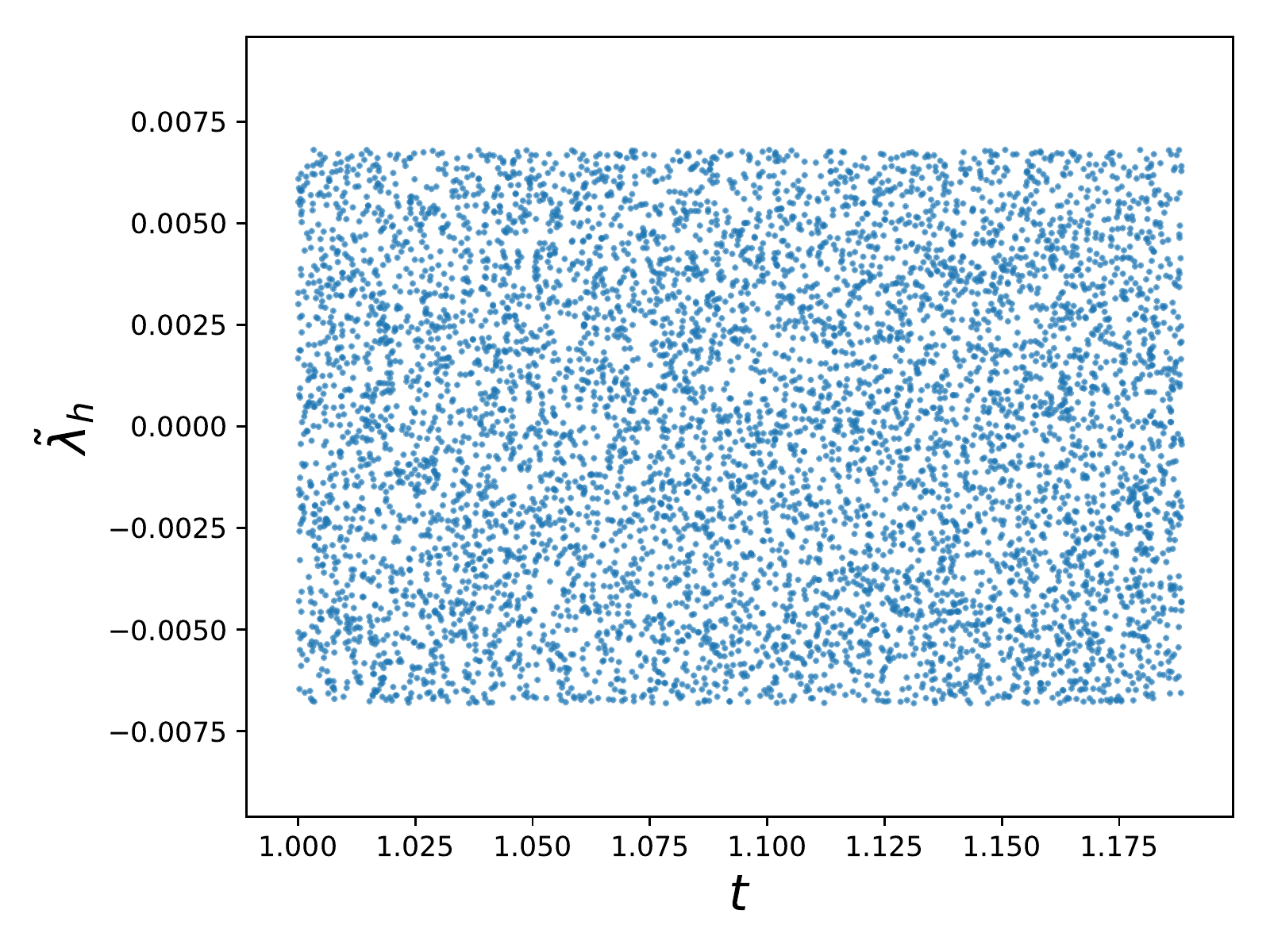}}
\caption{$t$ vs. $\tilde \lambda_h$, showing that the scan uniformly fills the region defined by the constraints arising from the invisible widths.  }
\label{thcon}
\end{figure}

\begin{figure}
\centerline{\includegraphics[width=4.5in,angle=0]{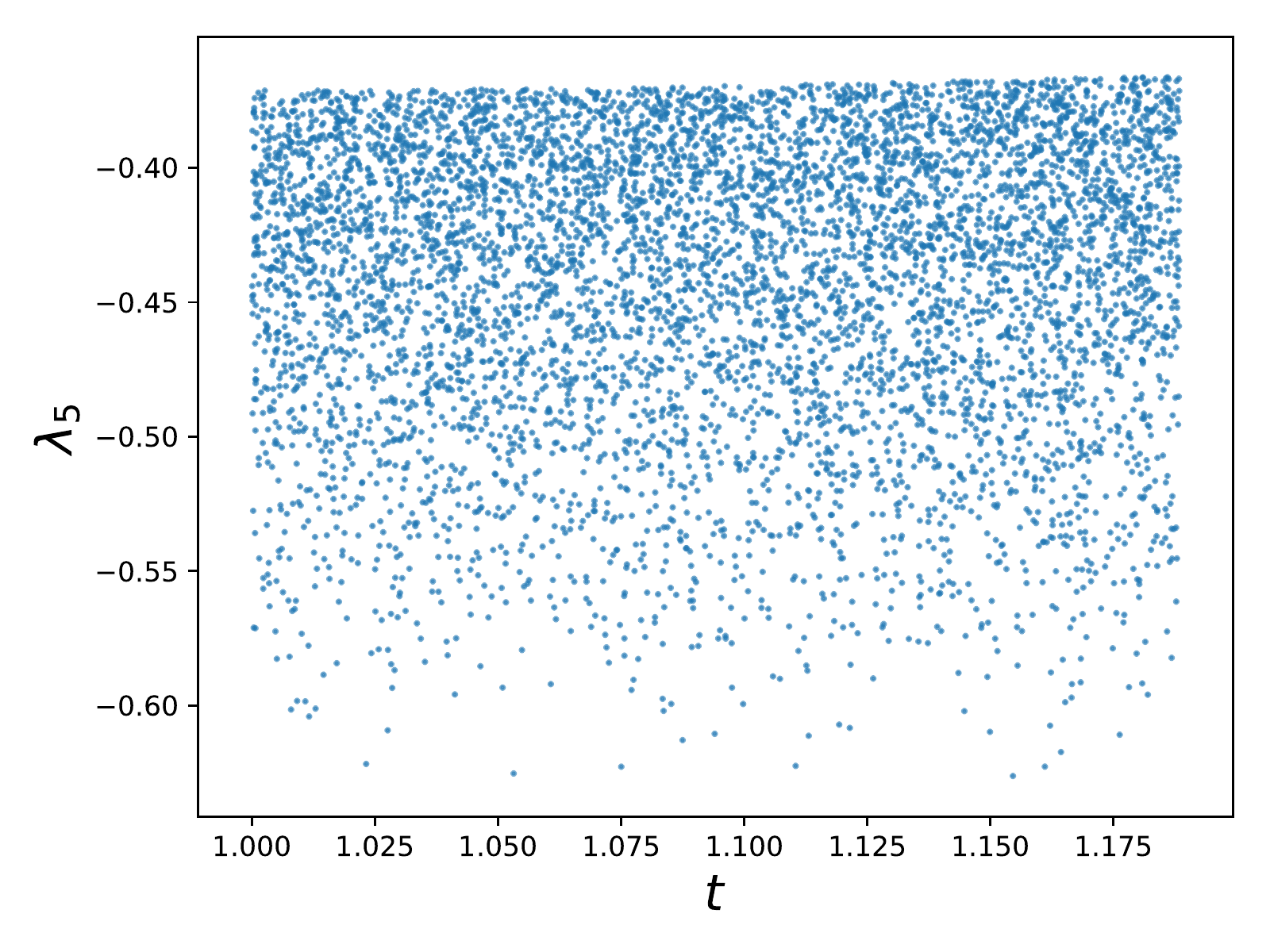}}
\caption{$t$ vs. $\lambda_5$, showing that smaller values of $\left| \lambda_5 \right|$, which correspond to smaller values of $m_A$ are preferred as in the previous Figure. The top boundary is defined by the $m_A > 150$ GeV bound, which forces $\lambda_5 < -2 (\frac{150\textrm{ GeV}}{v})^2 \frac{t}{1+t^2}$.  }
\label{tlam5}
\end{figure}

Fig. \ref{thcon} shows $\tilde \lambda_h$ vs. $t$. The scan clearly uniformly samples the allowed region of $t$ and $\tilde \lambda_h$ as defined by Eqs. \ref{tcon} and \ref{lamtilcon}, indicating that neither boundary of $\tilde \lambda_h$ or of $t$ is preferred by the constraints on masses and the $\lambda_i$ outlined above. Fig. \ref{tlam5} displays the values of $t$ vs.$\lambda_5$, showing that smaller values of $\left| \lambda_5 \right|$, which correspond to smaller values of $m_A\approx m_H$ are preferred by the scan. The top boundary is defined by the $m_A > 150$ GeV requirement, which forces $\lambda_5 < -2 (\frac{150\textrm{ GeV}}{v})^2 \frac{t}{1+t^2}$, up to $\mathcal{O}(x_i^2)$ terms. 

\begin{figure}
\vspace*{0.5cm}
\centerline{\includegraphics[width=3.2in,angle=0]{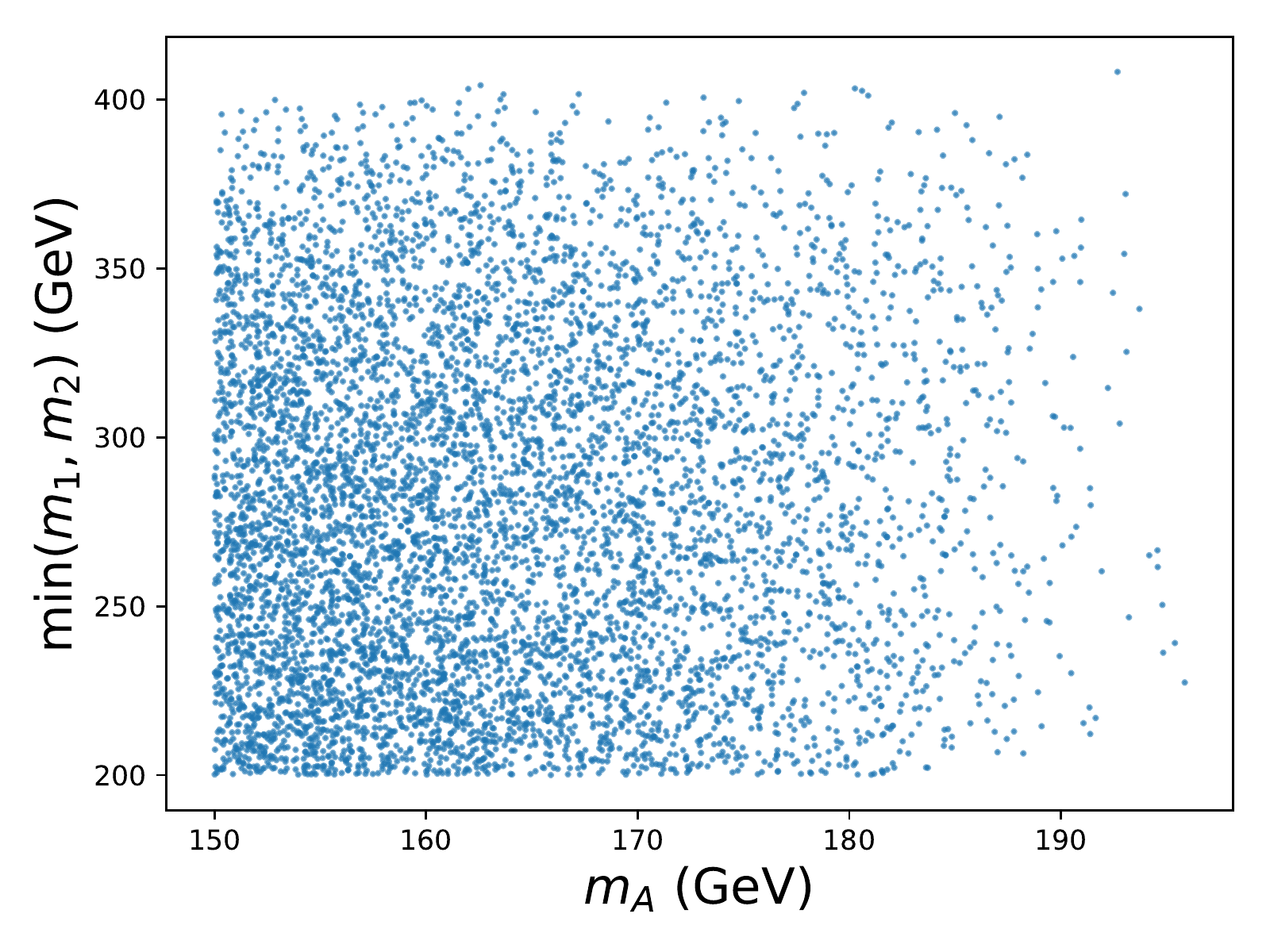}
\hspace*{-0.4cm}
\includegraphics[width=3.2in,angle=0]{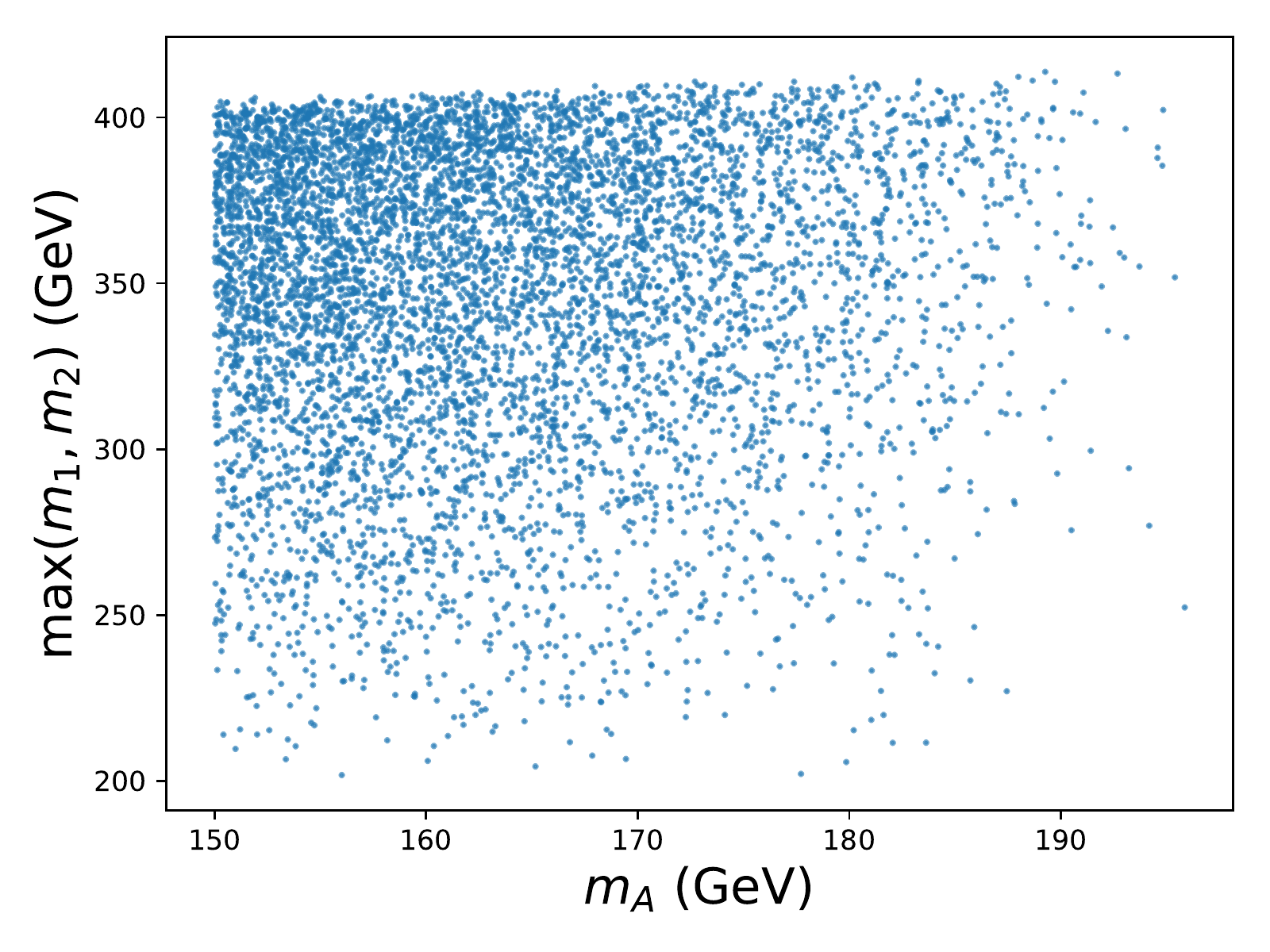}}
\caption{$m_A\approx m_H$ plotted against min($m_1,m_2$) (left) and max($m_1,m_2$) (right). The preference for smaller $m_A$ can be seen by the relative over density of points near 150 GeV, and we see that there is no comparable preference for low $m_{1,2}$, as the points are relatively uniform above the constraint $m_{1,2}>200$ GeV. The slight upward tilt of the boundary on the top of the right panel reflects the fact that increasing $m_A$ requires larger $-\lambda_5$, which increases both $m_1$ and $m_2$, even when $\left|\lambda_{4i}\right|$ is nearly maximal.}
\label{mam1m2}
\end{figure}

Fig. \ref{mam1m2} left and right shows $m_A\simeq m_H$ plotted against min($m_1,m_2$) and max($m_1,m_2$), respectively. The preference for smaller $\left| \lambda_5 \right|$, and thus smaller $m_A$, can be seen by the relative overdensity of points near 150 GeV, and we see that there is no comparable preference for low $m_{1,2}$, as the points are relatively uniform above the constraint $m_{1,2}>200$ GeV. The slight upward tilt on the top of the right panel reflects the fact that increasing $m_A$ requires larger $-\lambda_5$, which increases both $m_1$ and $m_2$ even when $\left|\lambda_{4i}\right|$ is nearly maximal.

\begin{figure}
\vspace*{0.5cm}
\centerline{\includegraphics[width=3.2in,angle=0]{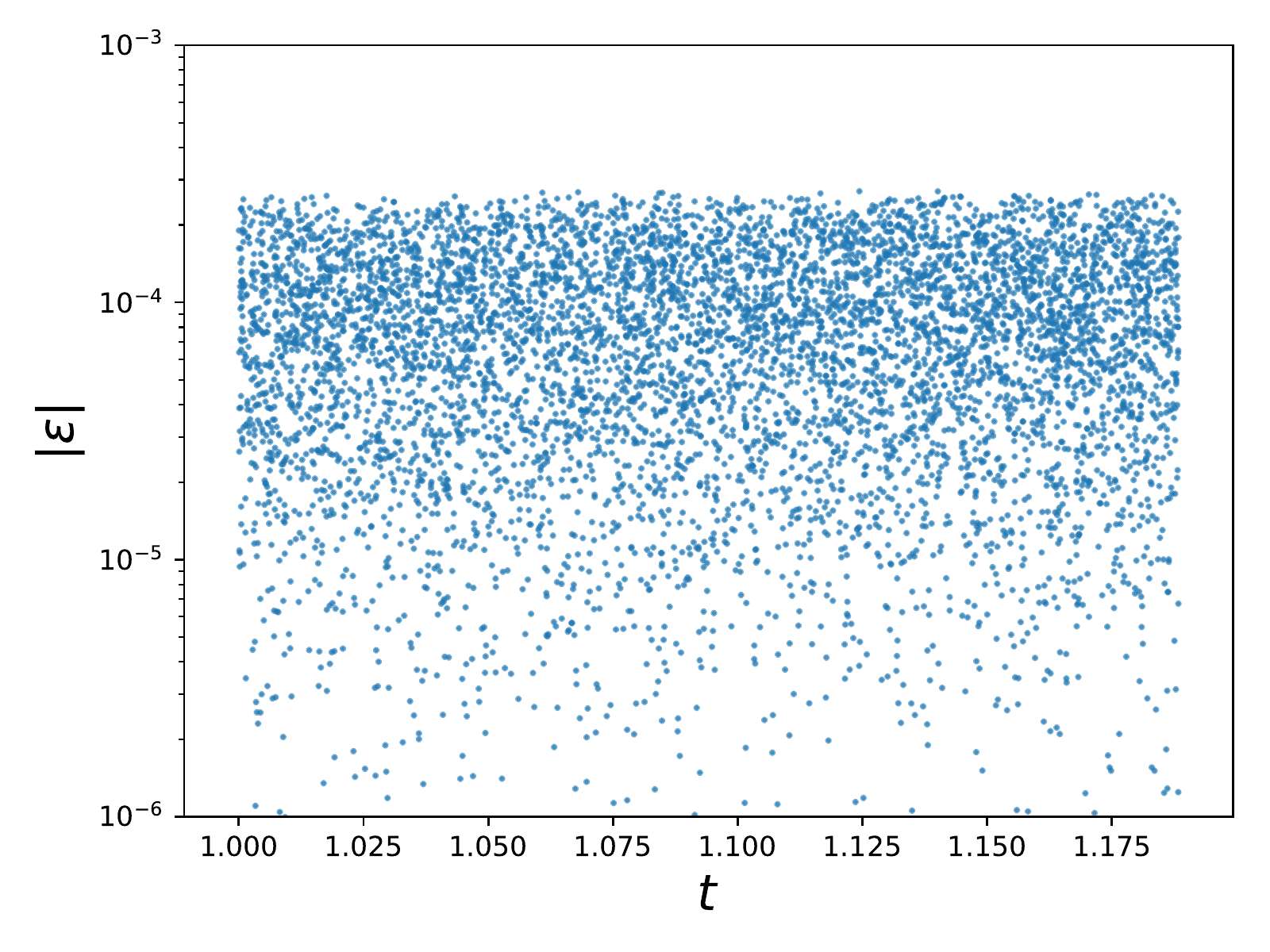}
\hspace*{-0.4cm}
\includegraphics[width=3.2in,angle=0]{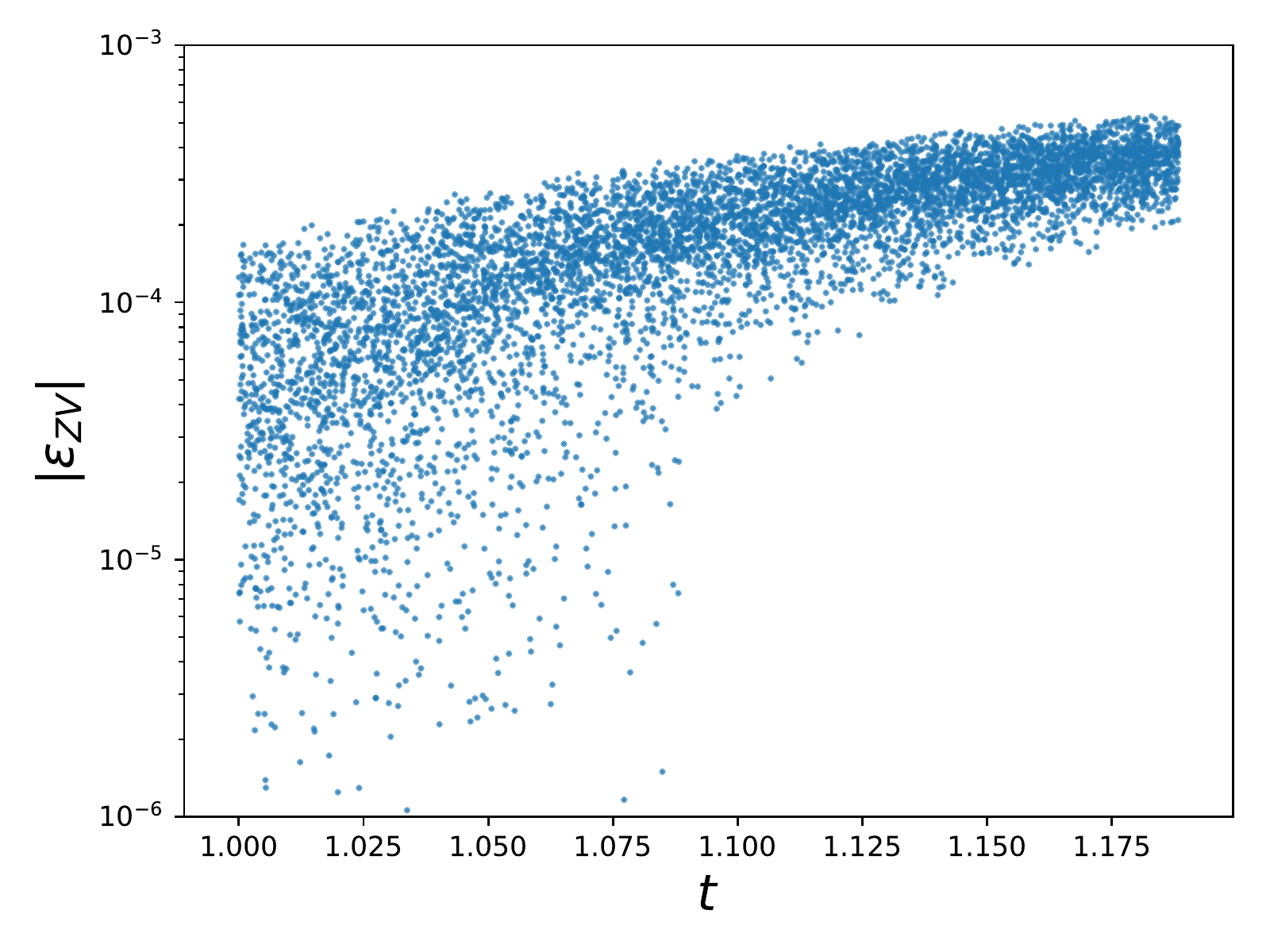}}
\caption{Left: $\left| \epsilon \right|$, up to an overall factor of $g_D/e$, plotted against $t$. Since $\epsilon$ depends linearly on $g_D$, the points in this plot will scale up or down linearly with a scaling of $g_D/e$. Right: $\left| \epsilon_{ZV} \right|$, taking $g_D = e$, plotted against $t$. Since $m_V$ depends on $g_D$, $\epsilon_{ZV}$ is only approximately linear in $g_D/e$, with a subleading dependence coming through the $m_V$ dependence of ln$\left( \frac{m_V^2}{m_A^2} \right)$ - $6 G(\delta)$. When $t=1$, $\epsilon_{ZV}$ is dominated by the contribution from the charged Higgs fields, and as $t$ increases $\epsilon_{ZV}$ comes to be dominated by the neutral sector contribution, which is proportional to $c_{2\theta} \simeq (t^2-1)/(1+t^2)$. }
\label{fig:eps}
\end{figure}

The kinetic mixing parameters $\epsilon$ and $\epsilon_{ZV}$, given by Eq. \ref{eq:eps} and Eq. \ref{eq:epszv}, are plotted against $t$ in Fig. \ref{fig:eps}, up to an overall factor of $g_D/e$. The value of $\epsilon$ will scale linearly with $g_D$, while $\epsilon_{ZV}$ scales approximately, though not exactly, linearly with $g_D$, as there is a subleading dependence on $g_D$ through the $m_V$ dependence of the neutral sector contribution $\frac{c_{2\theta}}{2}\left( \textrm{ln}\left( \frac{m_V^2}{m_A^2} \right) - 6 G(\delta) \right)$. The right panel of Fig. \ref{fig:eps} shows the $t$ dependence of $\epsilon_{ZV}$, which comes primarily through $c_{2\theta}\simeq (t^2-1)/(1+t^2)$. While $m_1$, $m_2$, and $m_A$ all depend on $t$, this logarithmic dependence is subleading from that arising from the $c_{2\theta}$ dependence. When $t=1$ the sole contribution to $\epsilon_{ZV}$ comes from the charged Higgs fields, and as $t$ increases the neutral sector contribution becomes increasingly important until it dominates $\epsilon_{ZV}$ near $t\approx 1.18$. 

\subsection{LHC Signals}

The LHC is capable of producing the various BSM Higgs fields through their couplings to the $W^\pm$, $\gamma$ and $Z$. The charged Higgs, $H_{1,2}^\pm$, can always decay into $W^\pm  h_d/V$ since the dark Higgs and dark photon are light, and may sometimes decay into $W^\pm  H/A$ if this channel is kinematically accessible. Approximating the dark Higgs and dark photon as massless, the width for $H_1^\pm$ and $H_2^\pm$ at leading order are given by 

\begin{equation} \label{gamh1}
\Gamma(H_1^\pm) \simeq \frac{m_1}{16\pi}\left[-\lambda_{41} - \frac{\lambda_5}{t} \right] \left[c_\theta^2 (1-r_W)^3 + \Theta(m_1 - m_H-m_W) s_\theta^2 (1 - 2 (r_W+r_H) +(r_H-r_W)^2)^{3/2} \right],
\end{equation}
\begin{equation} \label{gamh2}
\Gamma(H_2^\pm) \simeq \frac{m_2}{16 \pi}\left[-\lambda_{42} - \lambda_5 t \right] \left[s_\theta^2 (1-r_W)^3 + \Theta(m_2 - m_H-m_W) c_\theta^2 (1 - 2 (r_W+r_H) +(r_H-r_W)^2)^{3/2} \right],
\end{equation}
where $r_X = m_X^2/m_{1,2}^2$ in Eq. \ref{gamh1} and \ref{gamh2}, respectively, $\Theta(x)$ is the Heaviside function which is 1 for $x \geq 0$ and 0 otherwise; we have taken $m_H \simeq  m_A$, and $\theta$ is defined by Eq. \ref{theta}.

\begin{figure}
\centerline{\includegraphics[width=4.5in,angle=0]{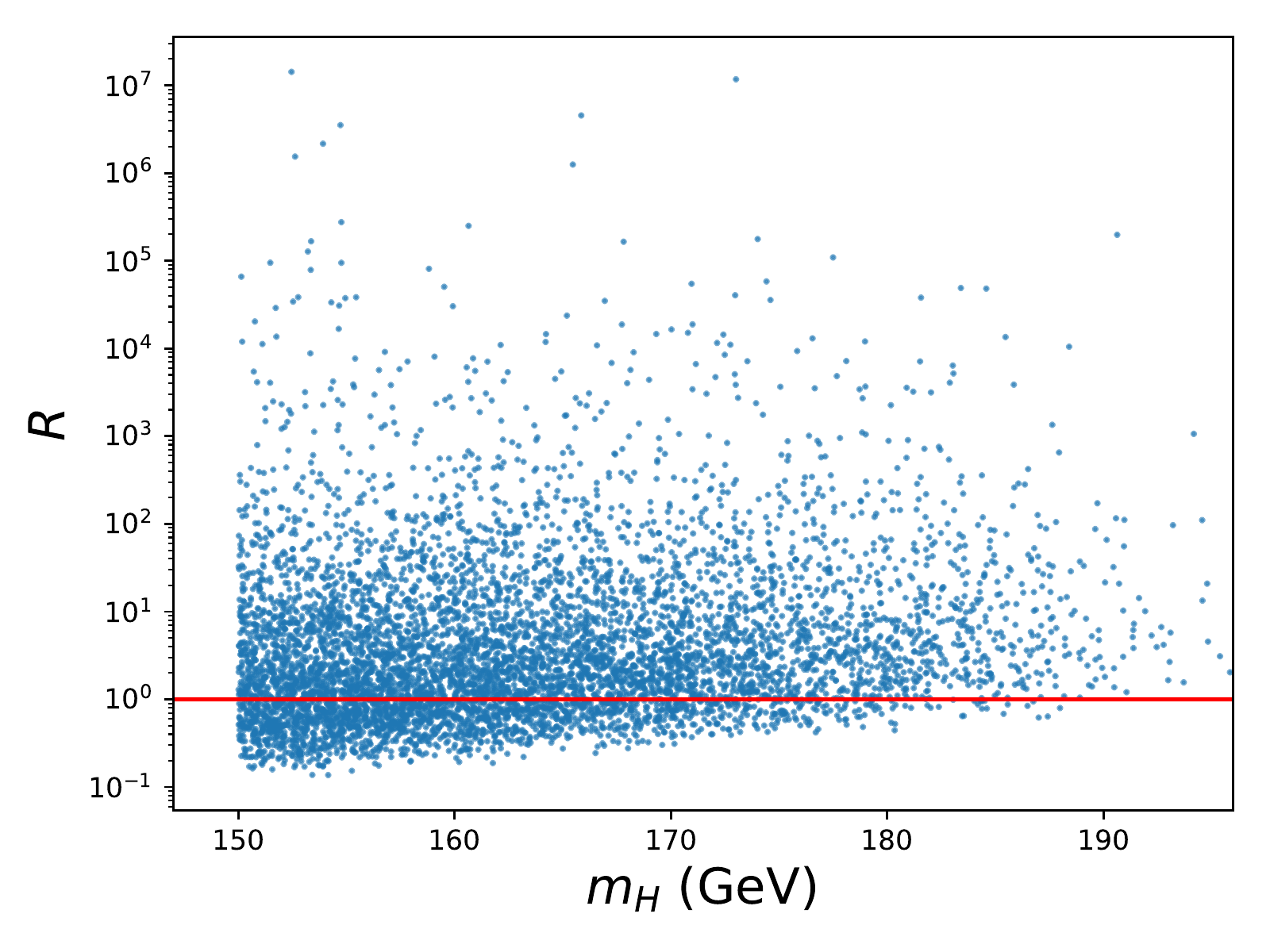}}
\caption{The ratio of partial decay widths $R$ plotted against $m_H$, showing that more points are $Z$-dominant than are $h_{SM}$-dominant, especially at large values of $m_H$. The red line corresponds to $R$=1 to guide the eye.}
\label{mhR}
\end{figure}

At leading order in the small parameters, $H$ decays are either as $H \rightarrow h_{SM} h_d$ or $H \rightarrow Z V$, and the corresponding $A$ decays are $A\rightarrow h_{SM} V$ and $A \rightarrow Z h_d$. Since the $H$ and $A$ form a neutral complex scalar up to $\mathcal{O}(x_i)$ effects, we can approximate $\Gamma(H\rightarrow h_{SM}+h_d) \simeq \Gamma(A\rightarrow h_{SM} V)$ and $\Gamma(H\rightarrow Z V) \simeq \Gamma(A\rightarrow Z h_d)$. At the same leading order in the small parameters these partial widths are given by

\begin{equation}
\Gamma(H\rightarrow h_{SM}h_d) = \frac{m_H s_{2\theta} }{64 \pi \lambda_5} [1-r_h][(\lambda_{31} + \lambda_{41} - \lambda_{32} - \lambda_{42}) s_{2\theta} +2 c_{2\theta} \lambda_5]^2 \equiv \frac{m_H s_{2\theta} \tilde \lambda^2}{64 \pi \lambda_5} [1-r_h],
\end{equation}
\begin{equation}
\Gamma(H\rightarrow ZV) = \frac{m_H s_{2\theta} \lambda_5}{16 \pi} [1-r_Z]^3 ,
\end{equation}
where $r_X = m_X^2/m_H^2$, and $\theta$ is defined by Eq. \ref{theta}. The ratio $R = \Gamma(H\rightarrow ZV)/\Gamma(H\rightarrow h_{SM} h_d)$ determines which decay mode is dominant, and thus what final states should be searched for at colliders. Fig. \ref{mhR} shows $R$ plotted against $m_{H}$, and we see that for $\simeq 72\%$ of the points in the parameter scan $H/A \rightarrow Z+V/h_d$ is the dominant decay mode. We also see from the Figure that for $m_H \gtrsim 175$ GeV nearly all the parameter space points lead to $R>1$, so that the decay into $H \to Z+ V/h_d$ dominates.

The  dark Higgs, $h_d$, will essentially only decay into $VV$ since its mixing with the SM Higgs is governed by sin($\theta_1$), which is suppressed by the constraint on $\tilde \lambda_h$, so that decays into light SM fermions are doubly suppressed by both the light fermion Yukawas as well as by sin$^2(\theta_1)$. Interestingly, we note in passing that 
$h_d$ remains a reasonably narrow state, $\Gamma(h_d)/m_{h_d} \simeq 5.5-6.0\%$, for the parameter space under study. 
We expect that $V$ will either completely escape a detector at the LHC, or perhaps to decay inside the calorimeter and produce a lepton jet. This will depend on its mass, the value of 
$\epsilon$ and the boost it experiences from its production from the decay of a heavier state.  
However,  to leverage current LHC searches it is most convenient to assume that $V$ almost always produces missing $E_T$ (MET), so that the decay signatures of interest are $H_{1,2}^\pm \rightarrow W^\pm+$MET and $H/A \rightarrow h_{SM}/Z$+MET depending on whether we have $R<1$ or $R>1$, respectively. We refer to points in parameter space with $R>1$ as ``$Z$-dominant", and those with $R<1$ as ``$h_{SM}$-dominant" in the following discussion.

Since the BSM Higgs fields only couple to the light fermions through their $\mathcal{O}(x_i)$ mixings with the SM Higgs, these new particles will predominantly be produced through the SM electroweak bosons in the $s$-channel at the LHC. In order to broadly probe the parameter space of this model, we select four benchmark points which roughly span the range of masses produced by the scan over parameters and the possible final states. The four benchmarks may be categorized by the masses of the $H/A$, the masses of $H_{1,2}^\pm$, and the dominant decay mode of the $H/A$ (either to $Z$+MET or to $h_{SM}$+MET). These are summarized in Table \ref{tab:BP}, and the full set of parameter values for each benchmark point (BP) are in given in the Appendix \ref{app:bps}. BP1 and BP2 are both $Z$-dominant, with BP1 featuring $H/A,$ $H_1^\pm$, and $H_2^\pm$ being on the heavier end of the scanned space, while BP2 has relatively light  $H/A,$ $H_1^\pm$, and $H_2^\pm$. BP3 and BP4 are $h_{SM}$-dominant, with BP3 having BSM Higgs masses on the heavier end of the scan range, and BP4 featuring BSM Higgs masses on the lighter end of the scan range.

\begin{table}[b] 
\begin{center}
\begin{tabular}{ |c |c |c| c| c|}
\hline
Benchmark Point & $m_H$ & $m_1$ & $m_2$ & $Z$ or $h_{SM}$ dominant\\
\hline
BP1 & 180.8 GeV & 371.0 GeV & 333.2 GeV & $Z$ \\
BP2 & 154.7 GeV & 203.9 GeV & 249.0 GeV & $Z$ \\
BP3 & 187.8 GeV & 305.6 GeV & 346.2 GeV & $h_{SM}$ \\
BP4 & 155.7 GeV & 210.5 GeV & 275.3 GeV & $h_{SM}$ \\
\hline
\end{tabular}
\end{center}
\caption{Four benchmark points and their mass parameter values used to analyze the efficiency of LHC searches for the model. These roughly span the range of $m_H$, $m_1$, and $m_2$ produced by the full parameter scan, with two $Z$-dominant points and two $h_{SM}$-dominant points.} \label{tab:BP}
\end{table}

In order to analyze various LHC searches for the BSM Higgs fields, we have used FeynRules \cite{alloul2014feynrules} to produce UFO files, which are passed to MadGraph5\_aMC@NLO \cite{alwall2014automated} to generate parton-level events. These events are showered with Pythia8 \cite{sjostrand2015introduction}, and DELPHES 3 \cite{de2014delphes} is used to simulate detector effects. Searches conducted by ATLAS (CMS) have used the default ATLAS (CMS) card without pileup effects and modified so that the $h_d$ and $V$ would not deposit energy in the calorimeters. Further modifications were made on a search-by-search basis, depending on the search parameters such as $b$-tagging efficiency, reconstruction efficiencies of various physics objects, or isolation cuts as stated in the searches. When searches did not state explicit cuts or procedures for isolation of physics objects, the default DELPHES 3 loose cut parameters were used. FastJet \cite{cacciari2011fastjet} was then used to reconstruct final state jets based on the jet algorithm stated in each search. 

\begin{figure}
\vspace*{0.5cm}
\centerline{\includegraphics{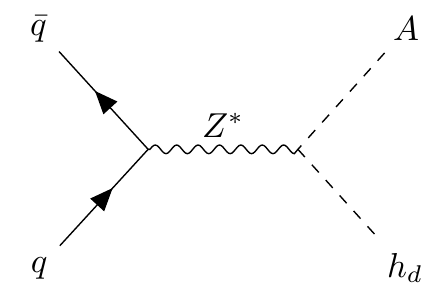}}
\caption{The dominant production processes for $A h_d$ at the LHC. A similar diagram, with $A \rightarrow H$ and $h_d \rightarrow V$, is responsible for $H V$ production. There are similar diagrams with $V^*$ in the $s$-channel, but these are suppressed by a factor of $\epsilon^2$. }
\label{haprod}
\end{figure}

The first set of LHC searches we consider are for $Z$+MET final states at the $\sqrt{s}=13$ TeV LHC, which should be sensitive to $Z$-dominant points in parameter space such as BP1 and BP2. The largest signal will come from associated production of $HV$ and $A h_d$ through the $Z$ in the $s$-channel, as shown in Fig. \ref{haprod}. with production determined entirely by measured SM quantities and the values of $m_{H,A}$. We note that there are also contributions from diagrams with $V$ in the $s$-channel, but these are suppressed by a factor of $\epsilon$ in the amplitude, and may be safely neglected. Since the $H/A$ in these events are always produced in association with a $V/h_d$ (which leads to MET), when they decay into $Z$+MET it is likely that a portion of the MET from the primary $V/h_d$ will be balanced by the MET resulting from the $V/h_d$ secondaries, thus reducing the overall event MET and lowering the event acceptance for searches with somewhat high cuts on the MET threshold. This will limit the LHC's capability to find these states, especially when they are relatively light.

For hadronically decaying $Z$ bosons, the best constraints arise from the $\sqrt{s}={13}$ TeV ATLAS search with $36.1\textrm{ fb}^{-1}$ of integrated luminosity \cite{ATLAS2018hadronic}. Since we expect our kinematics to differ significantly from the mono-$Z$ and mono-$Z'$ searches, which use the MET distribution to place constraints on simplified models, we rely on the model-independent limits on production cross section times acceptance times efficiency, denoted by $A_{eff}=( \mathcal{A} \times \textrm{efficiency})$, to search for our benchmark point models. The model independent limits on the visible cross section $\sigma_{\textrm{vis}}(\textrm{MET}) = \sigma_{Z+\textrm{MET}}(\textrm{MET}) \times Br(Z\rightarrow q \bar q) \times A_{eff}$(MET) are binned by MET, with $\sigma_{Z+\textrm{MET}}(\textrm{MET})$ and $A_{eff}$ (MET) both being functions of MET. Denoting the 95\% CL limit on the observed visible cross section by $\sigma_{\textrm{vis,lim}}$, we find that for BP1, which has $m_H \simeq 181$ GeV, the most sensitive search bin is MET$\in [400,600)$ GeV, with $\sigma_{\textrm{vis, lim}}/\sigma_{\textrm{vis, BP1}} \approx 13.4$. For the lighter case of $m_H\simeq 155$ GeV in BP2 we find that the MET$\in [200,250)$ GeV bin provides the strongest limit, with  $\sigma_{\textrm{vis, lim}}/\sigma_{\textrm{vis, BP2}} \approx 11.8$, while the second strongest constraint is from the MET$\in [400,600)$ GeV bin which leads to with  $\sigma_{\textrm{vis, lim}}/\sigma_{\textrm{vis, BP2}} \approx 12.2$. If improvements in the background suppression techniques and increased integrated luminosity can lead to stronger constraints by a factor of $\simeq 13$ or more, this search may be able to probe the $Z$-dominated parameter points of these models. 

When the $Z$ decays leptonically, searches again use MET distributions as discriminants to set limits on simplified models of dark matter, which we expect to differ significantly from the MET distributions generated by $HV$ and $A h_d$ associated production. We instead can approximate a probe the of $Z$-dominated parameter points by using the implied limits on $\sigma(Zh_{SM}\rightarrow l^+ l^- + \textrm{inv.})$ from the SM value of the $ZH$ production cross section and the corresponding reported limits on $B(h_{SM}\rightarrow \textrm{inv})$ in Refs. \cite{ATLAS2018leptonic,CMS2020leptonic}. Ref. \cite{ATLAS2018leptonic} uses 36.1 fb$^{-1}$ of data and reports a 95\% CL upper limit of 40 fb on $\sigma(Zh_{SM} \rightarrow l^+l^-+\textrm{inv.})$ and a 95\% CL upper limit of $Br(h_{SM}\rightarrow \textrm{inv.})<67$\%, which corresponds to $\sigma_{\textrm{BP1}}(Z+\textrm{MET}) \lesssim 530$ fb and $\sigma_{\textrm{BP2}}(Z+\textrm{MET}) \lesssim 740$ fb, after accounting for differences in $A_{eff}$ due to the event selection cuts. Similarly, Ref. \cite{CMS2020leptonic} uses 137 fb$^{-1}$ of data and reports a 95\% upper CL on $Br(h_{SM}\rightarrow \textrm{inv.})<29$\%, which translates into limits of $\sigma_{\textrm{BP1}}(Z+\textrm{MET}) \lesssim 294$ fb and $\sigma_{\textrm{BP2}}(Z+\textrm{MET}) \lesssim 380$ fb. Since the production cross sections for $Z$+MET at $\sqrt{s}=13$ TeV are $\sigma_{\textrm{BP1}}(Z+\textrm{MET})=226$ fb and $\sigma_{\textrm{BP2}}(Z+\textrm{MET})=403$ fb, we see that the search in Ref. \cite{CMS2020leptonic} might be able to probe part of our parameter space, though a more careful study than this naive estimate, ideally using the unique MET distribution expected in this model, is required to definitively rule out points in parameter space. The luminosity gains from the HL-LHC would seem to make future versions of these searches especially promising probes of the $Z$-dominant points of our model space. 

For $h_{SM}$-dominant points in the parameter space such as BP3 and BP4, we expect searches for $h_{SM}$+MET to be most sensitive. The largest signal for these searches will arise from the same associated production process as in the $Z$-dominant case, though now we expect the $H$ and $A$ to decay as $H/A\rightarrow h_{SM} + h_d/V$. Since the $h_d$ and $V$ will escape the detector and register as MET, we again expect the MET distributions to differ from simplified models which typically assume the MET is produced roughly back-to-back with the $h_{SM}$.

Model-independent limits exist for $h_{SM}+$MET final states with $h_{SM}\rightarrow b \bar b$ \cite{ATLAS2017hbb} and $h_{SM} \rightarrow \gamma \gamma$ \cite{ATLAS2017hgamgam}, both with 36.1 fb$^{-1}$ of data at the $\sqrt{s}=13$ TeV LHC. For the $h_{SM}\rightarrow b \bar b$ search, we assume that the $b$-jets are tagged at the 77\% efficiency working point of Ref. \cite{ATLAS2016btag}. The model-independent limits are set on $\sigma_{\textrm{vis}}$ and binned by MET, similarly to the hadronic $Z$+MET searches. We find that the MET$\in [200,350)$ bin is the closest to constraining both BP3 and BP4, with $\sigma_{\textrm{vis,lim}}/\sigma_{\textrm{vis,BP3}} \approx 9$ and $\sigma_{\textrm{vis,lim}}/\sigma_{\textrm{vis,BP4}} \approx 6$, so that an improvement by a factor of $\sim 10$ in this search would be capable of probing nearly all of the parameter space.  For the $h_{SM} \rightarrow \gamma \gamma$ search, there are four relevant categories which each have a model-independent limit on $\sigma_{\textrm{vis}}$. We find that for both BP3 and BP4 the Mono-Higgs event category is the most sensitive to the model, with $\sigma_{\textrm{vis,lim}}/\sigma_{\textrm{vis,BP3}} \approx 3.8$ and $\sigma_{\textrm{vis,lim}}/\sigma_{\textrm{vis,BP4}} \approx 2$. Since this 36.1 fb$^{-1}$ search is close to probing these benchmark points, we expect that the $h_{SM}\rightarrow \gamma \gamma$ search mode should be able to probe much of the $h_{SM}$-dominant parameter space with a factor of a few times more of integrated luminosity.

\begin{figure}
\vspace*{0.5cm}
\centerline{\includegraphics{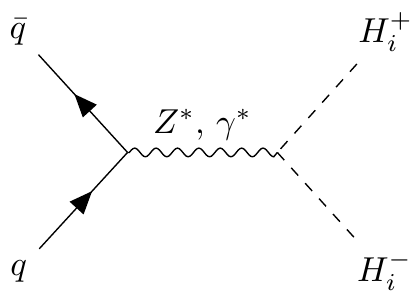}}
\caption{The dominant production processes for $H_i^+ H_i^-$ at the LHC. A similar diagram, with $V^*$ in the $s$-channel, is suppressed by a factor of $\epsilon^2$. }
\label{hpmprod}
\end{figure}

Searches for the charged states, $H_{1,2}^\pm$, rely on the $W^\pm+$MET in the final state, and should be sensitive to all of our benchmark points. Pair production of $H_{1,2}^\pm$ occurs primarily through $s$-channel $Z$ and $\gamma$ exchange, shown in Fig. \ref{hpmprod}, and the production rate is again dependent only upon the SM gauge couplings and the charged Higgs masses $m_{1,2}$. While diagrams with quarks in the $t$-channel also contribute due to mixing with the SM $H^\pm$, these contributions will be suppressed by $x_i^2$ in the amplitude, and are thus negligible. The $W^+ W^- +$MET final state produced by these events can be examined by using searches designed to look for chargino or slepton pair production. Leptonic decays of the $W^+ W^-$ provide the cleanest probe of these events, and ATLAS has performed such a search using 139 fb$^{-1}$ of data to place model-independent bounds on $W^+ W^-+$MET production in event categories binned by the stransverse mass, $m_{T2}$, of the leptons and by whether or not the leptons were same-flavor or different-flavor pairs \cite{ATLAS2020wmet}. These constraints prove quite insensitive to our benchmark points, with the closest bound still remaining a factor of $\sim 19$ above the prediction of BP4 in the different-flavor, 0-jet, $m_{T2} \in [120,160)$ bin. Generally to probe our benchmark points, the searches would need to improve their sensitivities by roughly factors of $\sim 70, 22,$ and $48$ to begin being sensitive to BP1, BP2, and BP3, respectively. We see that these limits are closer to probing the models with lighter $H_{1,2}^\pm$, due to the higher production cross sections 
but still remain rather far away.

\begin{figure}
\vspace*{0.5cm}
\centerline{\includegraphics{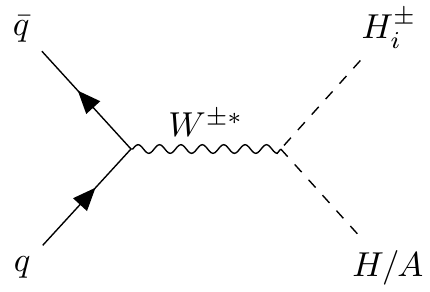}}
\caption{The dominant production processes for $H_i^\pm + H/A$ at the LHC. }
\label{hpmhaprod}
\end{figure}

The $H_{1,2}^\pm$ states may also be produced in association with $H/A$ via $W^\pm$ exchange in the $s$-channel, as shown in Fig. \ref{hpmhaprod}. The final state for this process depends upon whether the model is $Z$- or $h_{SM}$-dominant. For the $Z$-dominant cases, BP1 and BP2, searches for chargino/neutralino or slepton pair production with a $W^\pm Z+$MET final state are relevant, and ATLAS employed 36.1 fb$^{-1}$ of data to set model-independent limits on this process \cite{ATLAS2018wzmet}. We find the 2$l$+ jets and the 3$l$+0-jet searches to be the most sensitive to these benchmark points, with the SR2-int search region being the most sensitive to both BP1 and BP2. We find $\sigma_{\textrm{obs, SR2-int}}/\sigma_{\textrm{BP1, SR2-int}} \approx 8$, and $\sigma_{\textrm{obs, SR2-int}}/\sigma_{\textrm{BP2, SR2-int}} \approx 6$, so that with the higher integrated luminosity of the HL-LHC it may be possible to probe these benchmark points using the SR2-int search.

There are myriad other searches for charginos and neutralinos, which in principle may also be sensitive to the $Z$-dominant model benchmark points. Given the lower expected MET produced by events in this model, one may expect that the cleaner background at the lower energies of the 8 TeV LHC could perhaps better probe the parameter space. Two 20.3 fb$^{-1}$ ATLAS searches for electroweakinos and slepton pair production, with 3$l$+MET \cite{ATLAS8Twzmet3l} and 2$l$+MET \cite{ATLAS8Twzmet2l} final states, set relevant model-independent limits on the visible cross section, with lower cuts on MET than the 36.1 fb$^{-1}$ search of Ref. \cite{ATLAS2018wzmet}. The tightest limits across all bins are $\sigma_{\textrm{vis}} = 0.148$ fb for the SR0$\tau$a signal region of the 3$l$+MET search of $\sigma_{\textrm{vis}}=0.17$ fb. Calculating $\sigma \times BF(Z\rightarrow ll,W\rightarrow l\nu)\approx 0.053$ fb for BP1 and 0.297 fb for BP2, we see that the 3$l$ search is in principle only sensitive to BP2. After running the SR0$\tau$a search for BP2 we find that all bins are insensitive to the model. Ref. \cite{ATLAS8Twzmet3l} also performs searches for $W^\pm h_{SM}+$MET final states, but BP3 and BP4 do not produce large enough visible cross sections to be seen by these searches, even with perfect acceptance. Turning to the 2$l$+MET search, which looks for chargino and slepton pair production as well as chargino/neutralino production, we see that the chargino and slepton searches should be sensitive to $WW+$MET final states in this model. After calculating production cross sections times branching fractions for the various benchmark points, however, we see that none of them produce signals which would be visible in the SR-$m_{T2}$ or SR-$WW$ searches, even with perfect acceptance. The SR-$Z$jets search could be sensitive to the $Z$-dominant model point BP2, as BP1 again doesn't produce enough visible cross section even assuming perfect acceptance, but after performing the analysis we find that it is insensitive to the benchmark models considered here.

Additional searches for charginos, neutralinos, and sleptons have been performed using 139 fb$^{-1}$ of data at the 13 TeV LHC, and may also probe BP1 and BP2. However, an ATLAS search for compressed SUSY spectra in the 2$l$+MET final state \cite{ATLAS139wzmetcomp} will not be sensitive to $WZ+$MET final states produced here, since the search assumes off-shell $Z$ and $W$ in the decays and our model produces them on-shell. However, the slepton search is in principle sensitive to the $WW+$MET final states produced by our benchmark points. This search assumes that the sleptons recoil against a hard ISR jet, so we generate $W^+W^-j$+MET final states at 13 TeV for each of the four benchmark points. After making the $p_T$ cut on the leading jet $p_{T,j1} \geq 100$ GeV and multiplying by the branching fractions for the $W$'s to decay into same flavor lepton pairs, the visible cross sections are all found to be already below the lowest limit in the slepton search, thus rendering it insensitive to our model points. Another SUSY-inspired ATLAS search for 3$l$+MET final states \cite{ATLAS139wzmet3l} with 139 fb$^{-1}$ sets model-independent limits on $WZ+$MET and $WZj+$MET final states which may be sensitive to BP1 and BP2. After performing this analysis, we find that the limits are still quite far away from probing our benchmark points. The closest bound for BP1 comes from the SR-low search region, which has $\sigma_{\textrm{vis,lim}}/\sigma_{\textrm{vis,BP1}} \approx 170$. The nearest probe of BP2 is the SR-ISR signal region, which has $\sigma_{\textrm{vis,lim}}/\sigma_{\textrm{vis,BP2}} \approx 40$. These searches are unlikely to be sensitive to these benchmarks at the HL-LHC from the additional luminosity alone but would require substantial analysis improvements.

Associated production of $H_{1,2}^\pm$ with $H/A$ can be probed with searches for $W h_{SM}+$MET in the $h_{SM}$-dominant points of parameter space, with $h_{SM} \rightarrow b \bar b$ \cite{ATLAS2020whmetbb} or $h_{SM} \rightarrow \gamma \gamma$ \cite{ATLASwhmetgamgam}. The model-independent limits of Ref. \cite{ATLAS2020whmetbb} come from the channel  $W(\rightarrow l \nu) h_{SM}(\rightarrow b \bar b)+$MET in 139 fb$^{-1}$ of data, and require MET $> 240$ GeV for event selection. This is a high threshold for our model's events to pass, since both $H_{1,2}^\pm$ and $H/A$ will produce MET in their decays which will tend to somewhat balance one another to some degree, thus lowering the overall event MET. We find the most sensitive constraint in this case arises from the SR-LM model-independent search, though we find $\sigma_{\textrm{obs, SR-LM}}/\sigma_{\textrm{BP3, SR-LM}} \approx 28$ and $\sigma_{\textrm{obs, SR-LM}}/\sigma_{\textrm{BP4, SR-LM}} \approx 20$. Thus the  sensitivity of this search must improve by a factor of 20-30 to probe these models in this channel. 

The model-independent limits on $W h_{SM}(\rightarrow \gamma \gamma)+$MET set by Ref. \cite{ATLASwhmetgamgam} from 139 fb$^{-1}$ of data prove much more sensitive to BP3 and BP4. The most sensitive category is the ``Rest" Category 12, which requires MET significance $S_{MET}  = E_T^{\textrm{miss}}/\sqrt{\Sigma E_T}>9$, no leptons, and no jets with dijet mass consistent with a hadronic $W$ decay if there are at least 2 jets. Since this search category uses $S_{MET}$ rather than a cut on MET, the low MET events produced by this model may pass event selection. Since this category essentially searches for $h_{SM}\rightarrow \gamma \gamma$+MET due to the requirement that there not be an observed leptonic or hadronic $W$ decay, it is more sensitive to $h_{SM}+$MET events produced by $H/A$ being produced in association with $V/h_d$ rather than to $H_{1,2}^\pm+ H/A$ associated production. We find that for the $H_{1,2}^\pm+ H/A$ associated production events, $\sigma_{\textrm{vis,lim}}/\sigma_{\textrm{vis,BP3,}H_{1,2}^\pm+ H/A} \approx 35$ and $\sigma_{\textrm{vis,lim}}/\sigma_{\textrm{vis,BP4,}H_{1,2}^\pm+ H/A} \approx 14$, while for the $H/A + V/h_d$ associated production events $\sigma_{\textrm{vis,lim}}/\sigma_{\textrm{vis,BP3,}H/A + V/h_d} \approx 2$ and $\sigma_{\textrm{vis,lim}}/\sigma_{\textrm{vis,BP4,}H/A + V/h_d} \approx 1.2$. At this level, full NLO effects become important, and a relatively modest $K$-factor could render this search sensitive to BP4. With additional statistics from the HL-LHC, this search should be able to probe the $h_{SM}$-dominant points of our parameter space.

Table \ref{tab:searches} summarizes the factors by which various searches must improve in their sensitivities in order to probe the BP1-4 benchmark points. We emphasize that the production cross sections of the BSM Higgs fields in these models are governed entirely by SM couplings and the new scalar masses. This implies that any search which is sensitive to both $Z$- or $h_{SM}$-dominant BPs should be sensitive to the most, if not the entire, $Z$- or $h_{SM}$-dominant parameter space, since the BPs were chosen to roughly span the range of BSM masses generated by the scan. Several searches in combination could be sufficiently sensitive to probe the entire parameter space considered here with the statistics gained from the HL-LHC. In particular, searches targeted towards lower MET requirements can perform especially well in probing this model space since the production event topologies reduce the amount of observed MET relative to the back-to-back SM+MET topologies more typically targeted by, \eg, mono-searches. We note that the present searches are generally more sensitive to $H/A$ production signal events than to $H_{1,2}^\pm$ production events due to the lighter masses of the $H/A$ and the relative difficulty of reconstructing $W^\pm$ in the decays compared to the $Z$ or $h_{SM}$ decay products since these lead to invariant mass peaks whereas clean $W$ identification requires a leptonic decay which already contains MET.  

\begin{table}
\begin{center}
\begin{tabular}{ |c |c |c| c| c|}
\hline
Model & $Z(q \bar q)$+MET \cite{ATLAS2018hadronic} & $Z(l^+ l^-)$+MET \cite{CMS2020leptonic} & $h_{SM}(b \bar b)$+MET \cite{ATLAS2017hbb} & $h_{SM}(\gamma \gamma)$+MET \cite{ATLAS2017hgamgam} \\
\hline
BP1 & 13 & 1.3 & -- & -- \\
BP2 & 12 & 0.94 & -- & -- \\
BP3 & -- & -- & 9 & 3.8 \\
BP4 & -- & -- & 6 & 2 \\
\hline
 & $W^+W^-$+MET \cite{ATLAS2020wmet} & $WZ$+MET \cite{ATLAS2018wzmet} & $W h_{SM}$+MET  & $h_{SM}(\gamma \gamma)$+MET \cite{ATLASwhmetgamgam}\\
\hline
BP1 & 70 & 8 & -- & -- \\
BP2 & 22 & 6 & -- & -- \\
BP3 & 48 & -- & 28 ($b \bar b$)\cite{ATLAS2020whmetbb} & 2 \\
BP4 & 19 & -- & 14 ($\gamma \gamma$) \cite{ATLASwhmetgamgam} & 1.2 \\
\hline
\end{tabular}
\end{center}
\caption{The ratio $\sigma_{\textrm{vis,lim}}/\sigma_{\textrm{vis,BPx}}$ for the analysis bin providing the strongest constraint coming from the  various searches for the various final states produced in our scalar PM models at the LHC. Note that the limits for $Z \rightarrow l^+ l^-$+MET are estimates from searches for $Zh_{SM}\rightarrow l^+ l^- + \textrm{inv.}$ rather than model-independent limits. For final states with multiple searches, we display the result of the search with smallest value of $\sigma_{\textrm{vis,lim}}/\sigma_{\textrm{vis,BPx}}$. The $h_{SM}\rightarrow \gamma \gamma$+MET search in the lower half of the table reflects the Category 12 signal region of Ref. \cite{ATLASwhmetgamgam} applied to $H/A+V/h_d$ associated production events.}
\label{tab:searches}
\end{table}

\subsection{Probes through Rare Higgs decays} \label{sec:raredec}

The extended Higgs sector mediates additional decays with $h_d$ and $V$ in the final state, producing new contributions to $h_{SM}\rightarrow \gamma+$MET and $h_{SM}\rightarrow Z$+MET; we will discuss these modes in turn below.

\vspace{0.25cm}

($i$) $h_{SM} \to V\gamma$ 

\vspace{0.2cm}

\begin{figure}
\vspace*{0.5cm}
\centerline{\hspace{-1cm}\includegraphics{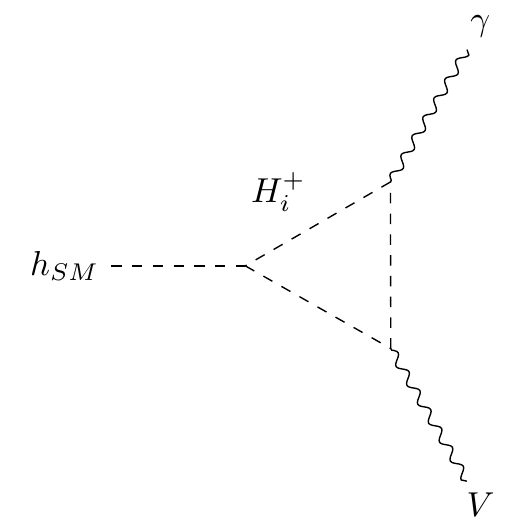} \hspace{1cm}
\includegraphics{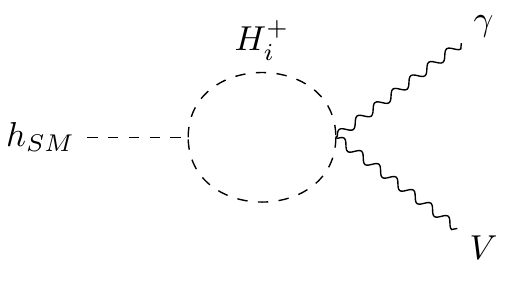}}
\caption{Diagrams contributing to $h_{SM} \rightarrow V \gamma$, which contribute to Br($h_{SM}\rightarrow \gamma$+inv.). These are mediated the charged Higgses $H_{1,2}^\pm$ through a triangle (left) and loop (right) diagram.}
\label{hVgamma}
\end{figure}

This reaction is the Scalar PM model analog of the SM $h_{SM}\to Z\gamma$ process and, as noted above, $V$ will likely appear as MET in the detector. In fact, the SM process with 
$Z\to \bar \nu \nu$ (which has a branching fraction of $\simeq 20\%$) provides the irreducible background for this reaction unless the photon energy in the Higgs rest frame can be 
determined. The corresponding LHC search where the $Z$ decays instead to 
$e^+e^-$ or $\mu^+\mu^-$ has recently been performed by ATLAS\cite{Aad:2020plj}; they obtain an upper limit of $B(h_{SM}\to Z\gamma) <5.5\cdot 10^{-3}$ for the relevant branching 
fraction. This result is roughly $\sim 3.6$ times greater than that of the SM prediction under the assumption that the Higgs production cross section is given by its SM value. A similar set of 
assumptions would then tell us that this bound implies the corresponding limit of $B(h_{SM}\to \gamma +invisible) < 1.1\cdot 10^{-3}$ would be expected by just employing the known SM 
branching fractions of the $Z$ into charged leptons and neutrinos; the actual SM prediction itself for the process $h_{SM}\to Z\gamma, Z\to \bar \nu \nu$ is $\simeq 3\cdot 10^{-4}$.

\begin{figure}
\centerline{\includegraphics[width=4.5in,angle=0]{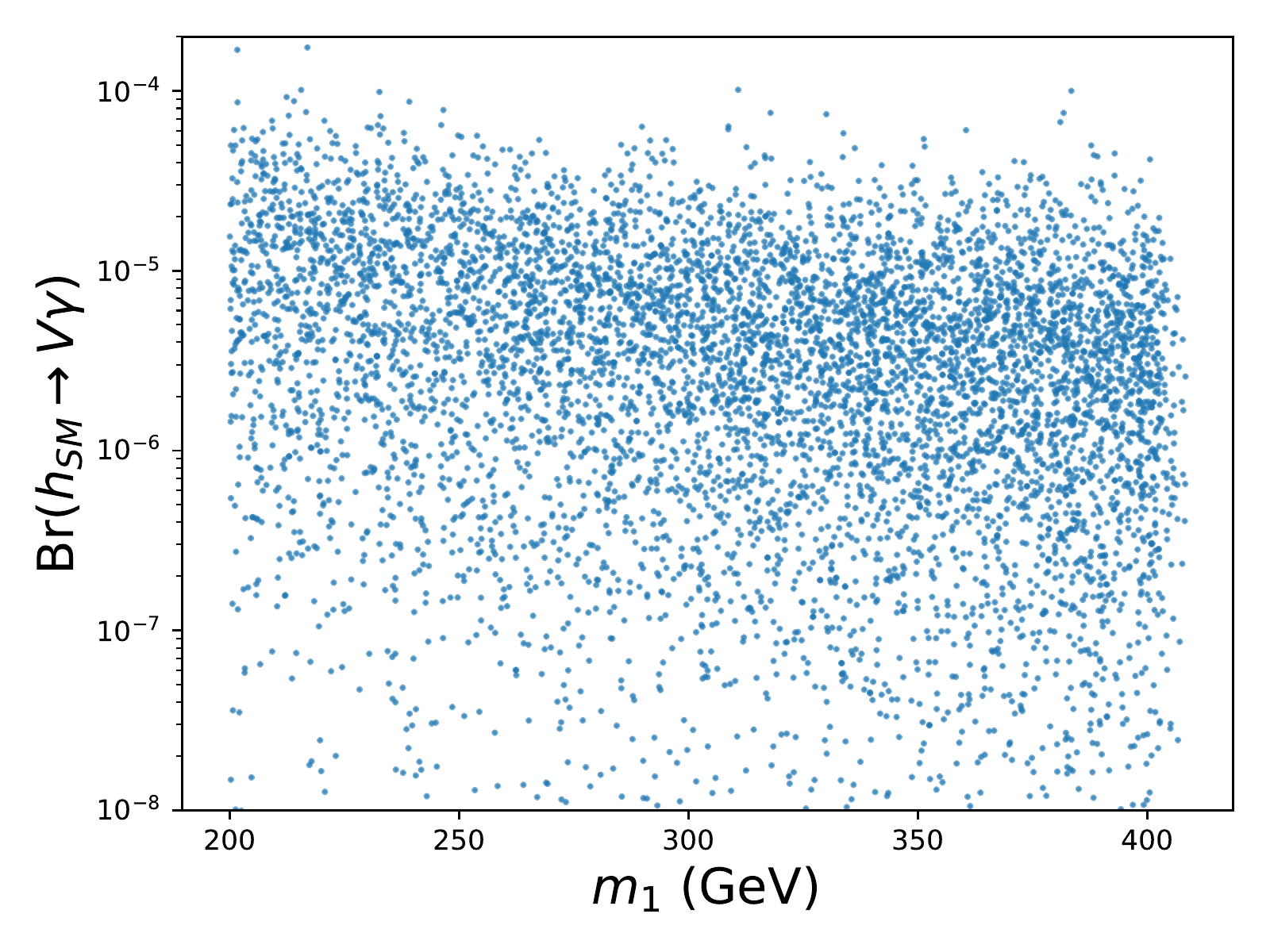}}
\caption{ Br$(h_{SM} \rightarrow V \gamma)$, up to an overall factor of $\frac{g_D^2}{e^2}$, vs $m_1$. Larger $m_1$ generally decreases the branching fraction of $h_{SM} \rightarrow V \gamma$, as expected from the behavior of $F$ in Eq. \ref{feq}.}
\label{brhvgam}
\end{figure}

In the Scalar PM model, the $h_{SM}\to V+\gamma$ process is the result of triangle and loop graphs, shown in Fig. \ref{hVgamma}, involving those PM fields which couple to $h_{SM}$ and carry both 
electric as well as dark charges, \ie, $H_{1,2}^\pm$ with dark charges $Q_D^i=\pm 1$. The partial width for this process can be written as
\begin{equation}
\Gamma(h_{SM}\to V\gamma)= \frac{m_h^3}{32\pi} ~\Big(1-\frac{m_V^2}{m_h^2}\Big)^3~ |A_{tot}|^2,
\end{equation}
where $A_{tot}$ is the total amplitude resulting from the sum of both $H_{1,2}^\pm$ loops which we can write in the form
\begin{equation}
A_{tot}= \frac{2\alpha}{\pi}~\frac{g_D}{e}~\frac{v}{m_h^2}~\sum_i ~Q_D^i c_i ~\frac{I_1(\tau_i,\mu_i)}{\tau_i},
\end{equation}
where we have defined the $h_{SM}H_i^+H_i^-$ coupling to be $c_i v$, and whose values can be read off from the set of couplings given above, $\mu_i=4m_{H_i^\pm}^2/m_V^2$, 
$\tau_i=4m_{H_i^\pm}^2/m_h^2$ and $I_1$ is the well-known function as given in, \eg, the Higgs Hunters Guide{\cite{Gunion:1989we}}{\footnote{See Eq.(2.24) on p.29.}}. 
Note that since $m_V \lsim 1$ GeV, we see that the $\mu_i>>1$ while $\tau_i \sim 10$ given the typical set of masses we have encountered above. Denoting this sum by 
$\Sigma$, we then find that 
\begin{equation}
\Gamma(h_{SM}\to V\gamma)\simeq \frac{m_h \alpha^2}{2\pi^3} ~\frac{g_D^2}{e^2}~ |\Sigma|^2,
\end{equation}
so that employing $\Gamma(h_{SM})=4.07$ MeV , $m_h=125.1$ GeV, $v=246.2$ GeV, and $\alpha^{-1}=127.935$ we obtain, numerically, that  
\begin{equation}
B(h_{SM}\to V\gamma)\simeq 0.233 ~\frac{g_D^2}{e^2}~ |\Sigma|^2.
\end{equation}
Now as $\mu_i \to \infty$, which is a reasonable numerical approximation here, $I_1$ becomes a function of just $\tau$ alone and we find that in this same limit
\begin{equation} \label{feq}
\frac{I_1}{\tau} \to -\frac{1}{2} \Big(1-\tau[\sin^{-1} (1/\sqrt{\tau})]^2\Big)=F(\tau) \simeq \frac{1}{6\tau} .
\end{equation}
where we have assumed $\tau$ is also large in the last step. 
\begin{figure}[htbp]
\centerline{\includegraphics[width=5.0in,angle=0]{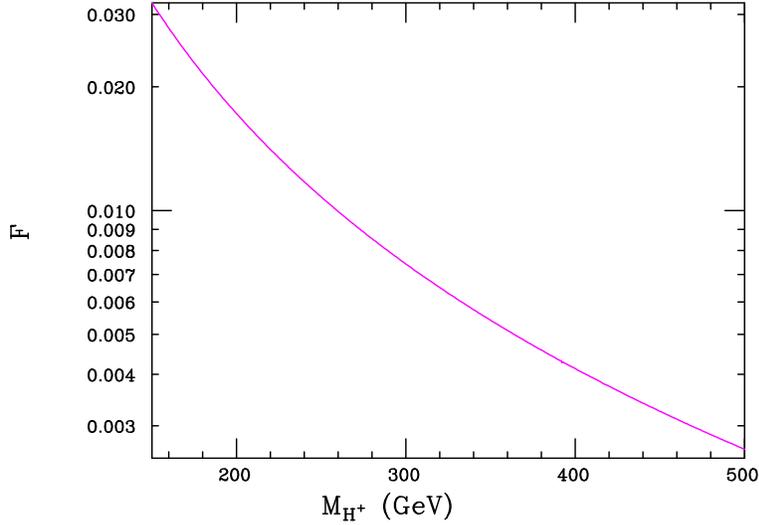}}
\vspace*{-1.30cm}
\caption{The quantity $F$, as defined in the text, as a function of the charged Higgs mass in the loop.}
\label{hvg}
\end{figure}
Fig.~\ref{hvg} shows $F$ as a function of the charged Higgs mass in the loop and we see that for our range of masses typical values $F\sim 0.01$ might be expected. With 
$\Sigma =c_1F_1-c_2F_2$ and the $c_i\sim O(1)$,  one might then anticipate a branching fraction of $B(h_{SM}\to V\gamma)\sim 10^{-5}$, barring cancellations, for the typical 
models in our scan, thus lying roughly a factor of $\sim 10-30$ below the SM predicted background. Fig.~\ref{brhvgam} shows that the bulk of the model points do indeed satisfy 
these expectations but also that cancellations between the two contributions can be quite important since the charged Higgs masses are generally not very different. 

\vspace{0.5cm}

($ii$) $h_{SM} \to Z+$MET 
\vspace{0.2cm} 

\begin{figure}
\vspace*{0.5cm}
\centerline{\hspace{-.5cm}\includegraphics[scale=0.9]{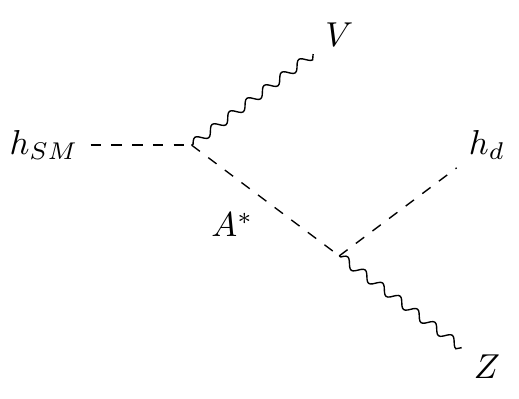} \hspace{.5cm}
\includegraphics[scale=0.9]{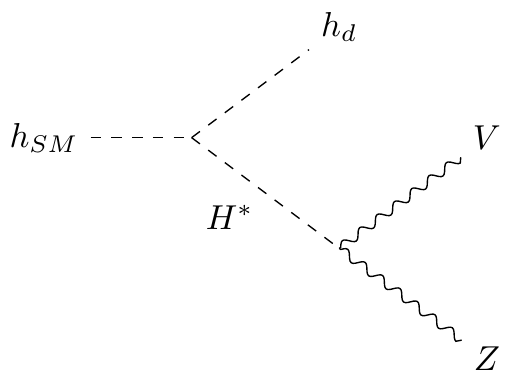}\hspace{.5cm}
\includegraphics[scale=0.9]{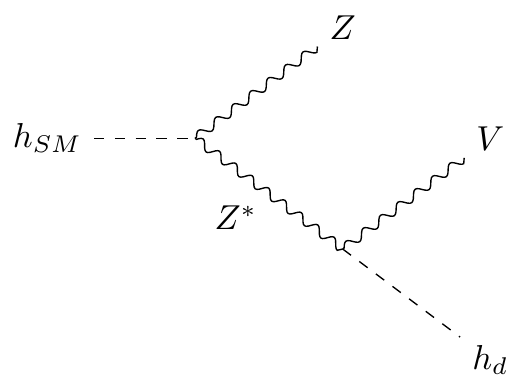}}
\caption{Three body decays of $h_{SM} \rightarrow Z V h_d$ which contribute to Br($h_{SM}\rightarrow Z$+inv.). These are mediated by a virtual $A^*$ (left), which gives rise to the amplitude $\mathcal{M}_1$ in the text, a virtual $H^*$ (center), which gives rise to $\mathcal{M}_2$ in the text, and a virtual $Z^*$ (right), which gives rise to $\mathcal{M}_3$ in the text.}
\label{h3inv}
\end{figure}

The $H$ and $A$ mediate additional decays which contribute to the $h_{SM} \rightarrow Z h_d V = Z+$MET decay mode at tree-level, with the relevant Feynman diagrams shown in Fig. \ref{h3inv}. There is also a contribution from a virtual $Z^*$, similar to the SM process for $h_{SM}\rightarrow Z \nu \bar \nu$, with an amplitude proportional to $c_{2\theta}$ which vanishes in the limit $t \rightarrow 1$. We define $\mathcal{M}_1$ to be the amplitude associated with the virtual $A$, $h_{SM} \rightarrow V A^*, A^*\rightarrow Z h_d$. We neglect the width of the $H$, $A$ and $Z$ since we are far off-shell, and assume that the coupling to the Goldstone $G_V^0$ represents the entire coupling to $V$, noting that the coupling to the transverse modes of $V$ will be suppressed by a factor of $x_i\simeq10^{-2}$. We find that

\begin{equation}
i \mathcal{M}_1 = \frac{\tilde \lambda m_Z s_{2\theta}}{2}\frac{(p_Z + 2p_{h_d})^\mu}{p_A^2 - m_A^2} \epsilon_s^*(p_Z)_\mu,
\end{equation} 
where $p_A^\mu = (p_Z+p_{h_d})^\mu$ and we have used $m_Z \simeq g v /(2 c_w)$. The second relevant amplitude, $\mathcal{M}_2$, is for the decay via a virtual $H$, $h_{sm} \rightarrow h_d H^*, H^*\rightarrow Z V$, and is given by

\begin{equation}
i \mathcal{M}_2 = -\frac{\tilde \lambda m_Z s_{2\theta}}{2}\frac{ (p_Z + 2p_V)^\mu}{p_H^2 - m_H^2} \epsilon_s^*(p_Z)_\mu,
\end{equation} 
where $p_H^\mu = (p_Z+p_V)^\mu$. Note that $\mathcal{M}_1$ and $\mathcal{M}_2$ have the same the same overall coupling coefficient, but will destructively interfere due to their relative signs. The third amplitude, $\mathcal{M}_3$, is for the decay via a virtual $Z$, $h_{SM} \rightarrow Z Z^*$, $Z^*\rightarrow h_d V$, and is given by

\begin{equation}
i \mathcal{M}_3 = \frac{g^2 m_Z c_{2\theta}}{2 c_w^2}\frac{ \left(g^{\mu \sigma} - \frac{p_{Z^*}^\mu p_{Z^*}^\sigma}{m_Z^2} \right)}{p_{Z^*}^2 - m_Z^2} (p_{h_d} - p_V)_\sigma \epsilon_s^*(p_Z)_\mu,
\end{equation}
where now $p_{Z^*}^\mu = (p_{h_d}+p_V)^\mu$. 

\begin{figure}
\vspace*{0.5cm}
\centerline{\includegraphics[width=3.2in,angle=0]{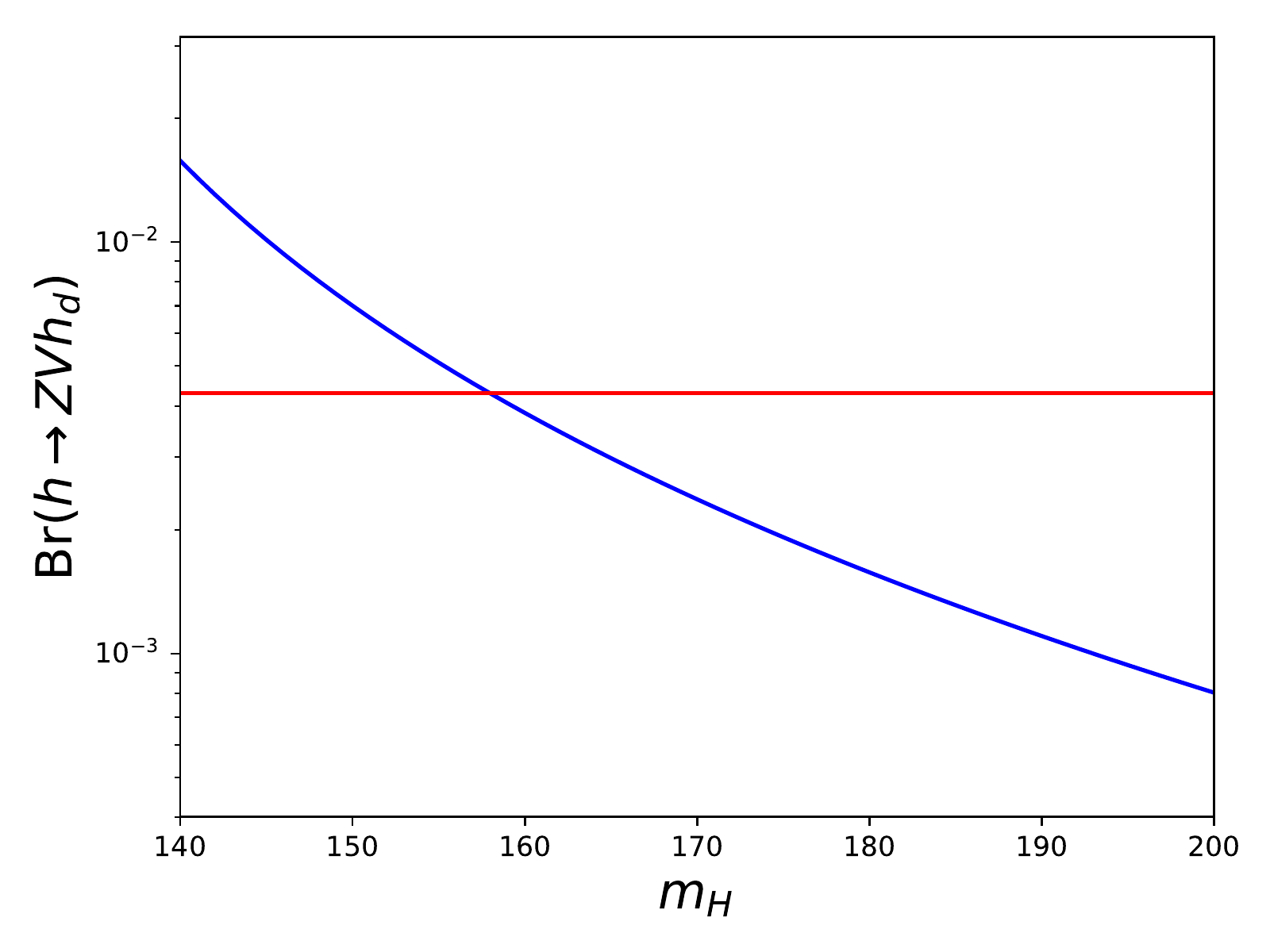}
\hspace*{-0.4cm}
\includegraphics[width=3.2in,angle=0]{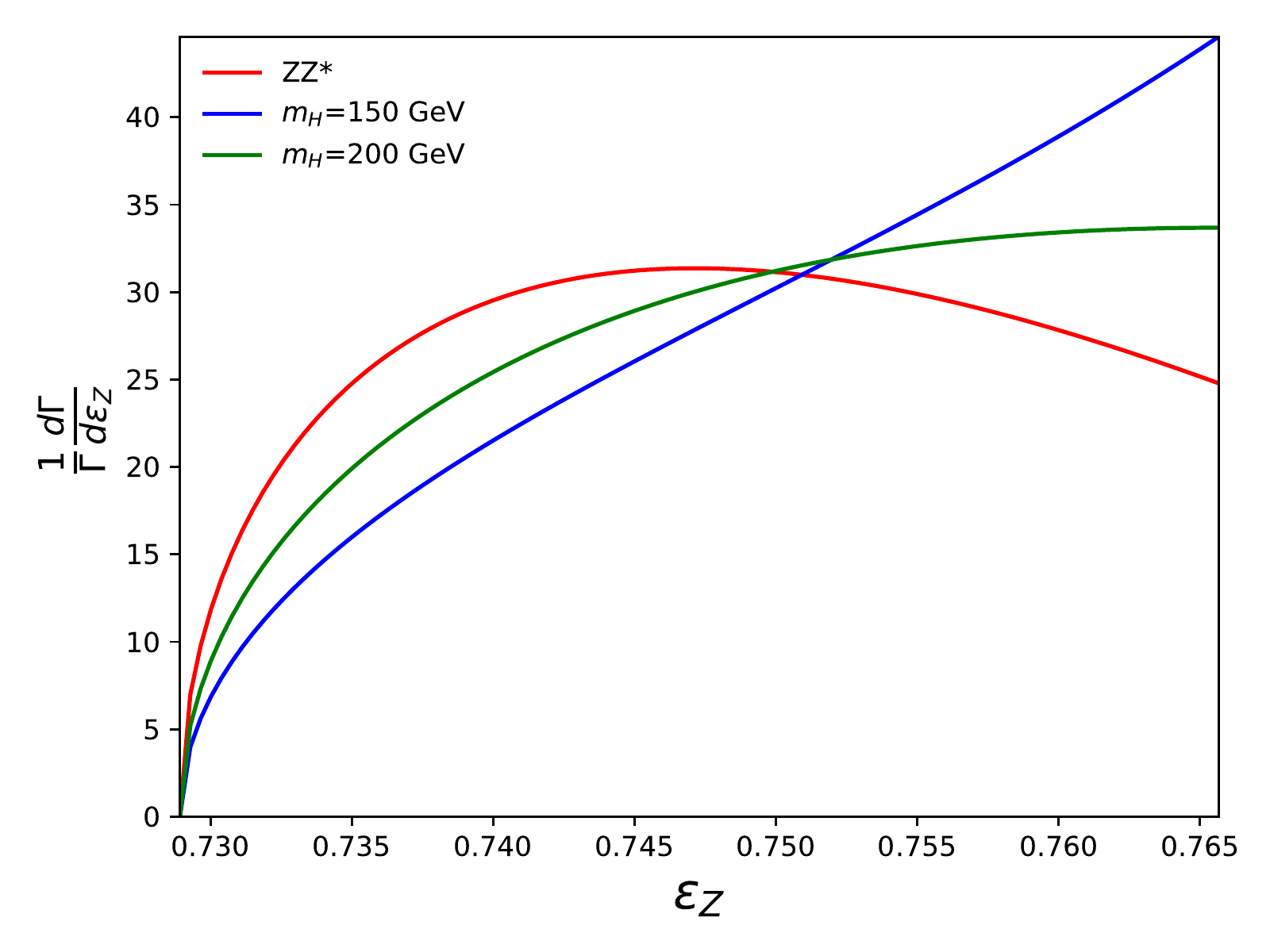}}
\caption{ Left: Branching fraction for $h_{SM}\rightarrow Z V h_d$ via virtual $H$ and $A$ (blue) and via $Z Z^*$ (red) vs. $m_H$, assuming the SM value of the Higgs width $\Gamma(h_{SM})=4.07$ MeV. Right: The normalized differential width for the SM (from $h_{SM}\rightarrow ZZ^* \rightarrow Z \nu \bar \nu$) (red) and BSM (from $h_{SM} \rightarrow Z V h_d$) channels, assuming $\tilde \lambda = t = 1$. For the BSM channels we show normalized differential widths for $m_{H}=150$ GeV (blue) and $m_H=200$ GeV (green), to fully cover the range of masses in the scan.}
\label{dGdez}
\end{figure}

The branching fraction for $h_{SM} \rightarrow Z V h_d$, assuming $\tilde \lambda = 1$ and $t=1$ so that $\mathcal{M}_3=0$, and taking the SM value $\Gamma(h_{SM})=4.07$ MeV , is shown in the left panel of Fig. \ref{dGdez} as a function of $m_H \approx m_A$. We see that in this case the BSM contribution to $h_{SM}\rightarrow Z+$MET from the $H^*$ and $A^*$ mediated decays exceeds the SM value of Br$(h_{SM} \rightarrow Z Z^*\rightarrow Z \nu \bar \nu)\approx 4.3 \times 10^{-3}$ for $m_H \lesssim 160$ GeV. The right panel of Fig. \ref{dGdez} shows the normalized differential width $(1/\Gamma) d \Gamma/d \epsilon_Z$ for the SM and BSM contributions to $h_{SM}\rightarrow Z$+MET, taking $\tilde \lambda = t = 1$ and where $\epsilon_Z = E_Z/m_{h_{SM}}$. This shows that the SM contribution dominates for $\epsilon_Z \lesssim 0.75$, while the BSM contribution dominates for $\epsilon_Z \gtrsim 0.75$, suggesting that the strength of the BSM contribution may be probed by a simple two bin analysis. For the benchmark points considered, the branching fractions for $h_{SM} \rightarrow Z V h_d$ and the ratios $\Gamma(h_{SM}\rightarrow Z+\textrm{MET}; \epsilon_Z > 0.75)/\Gamma(h_{SM}\rightarrow Z+\textrm{MET}; \epsilon_Z < 0.75)$, including both the SM and BSM contributions in the calculation of $\Gamma(h_{SM}\rightarrow Z+\textrm{MET})$, are listed in Table \ref{tab:BPratios}. We see that even for branching fractions of $h_{SM}\rightarrow ZVh_d$ smaller than the SM branching fraction, the enhancement of decays with $\epsilon_Z > 0.75$ relative to the SM case may be measurable, though of course increasing precision is required for smaller BSM branching fractions.

\begin{figure}
\vspace*{0.5cm}
\centerline{\includegraphics[width=5in,angle=0]{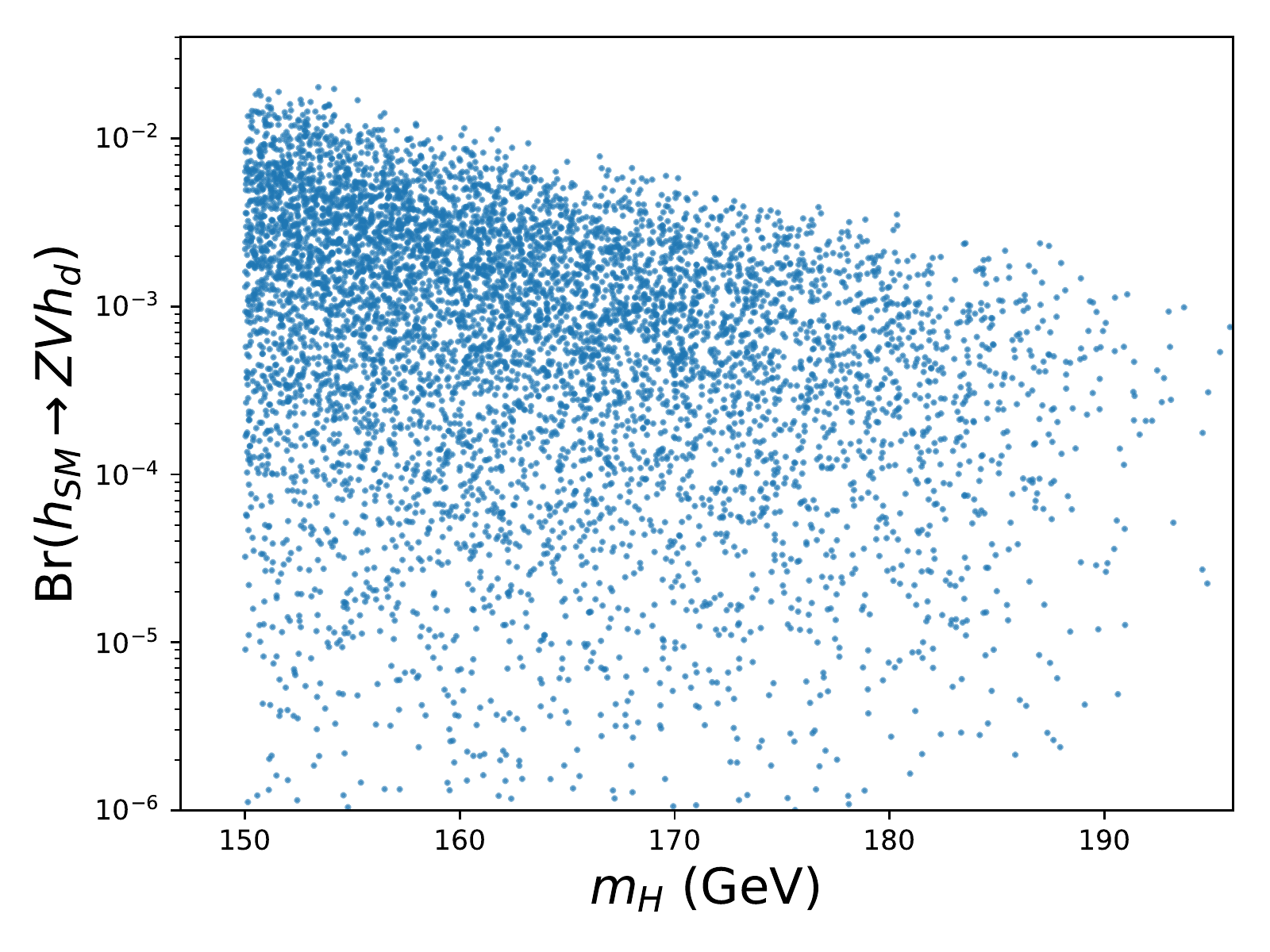}}
\caption{ Branching fraction for $h_{SM}\rightarrow Z V h_d$ vs. $m_H$, assuming the SM value of the Higgs width $\Gamma(h_{SM})=4.07$ MeV. While the majority of models have Br$(h_{SM}\rightarrow Z V h_d)\simeq0.01-1\%$, it is possible for the branching fraction to be quite suppressed due to small values of $\tilde \lambda$ and/or large destructive interference between the three decay amplitudes.}
\label{BrBSMscan}
\end{figure}

Fig. \ref{BrBSMscan} shows the branching fraction for $h_{SM}\rightarrow Z V h_d$ plotted against $m_H$ for the parameter space points from the scan; in general all three amplitudes 
will now contribute. We see that the branching fraction is typically $\sim 0.01-1\%$, but also that far smaller values are possible due to suppressed values of $\tilde \lambda$ and/or interference between the three contributing amplitudes. Expressions for $\Gamma(h_{SM} \rightarrow Z V h_d)$ and $ d\Gamma/d\epsilon_Z$ may be found in Appendix \ref{hzvhdcalc}. \begin{table} 
\begin{center}
\begin{tabular}{ |c |c |c |}
\hline
Benchmark Point & Br$(h_{SM} \rightarrow Z V h_d)$ & $\frac{\Gamma(h_{SM}\rightarrow Z+\textrm{MET}; \epsilon_Z > 0.75)}{\Gamma(h_{SM}\rightarrow Z+\textrm{MET}; \epsilon_Z < 0.75)}$ \\
\hline
SM & 0 & 0.807 \\
BP1 & $3.22\times 10^{-3}$ & 0.925 \\
BP2 & $1.87\times 10^{-4}$ & 0.828 \\
BP3 & $3.96\times 10^{-4}$ & 0.825 \\
BP4 & $1.41\times 10^{-5}$ & 0.809 \\
\hline
\end{tabular}
\end{center}
\caption{The branching fraction for $h_{SM}\rightarrow Z V h_d$ and the ratio of $h_{SM}\rightarrow Z +$MET events with $\epsilon_Z >0.75$ to $h_{SM}\rightarrow Z +$MET events with $\epsilon_Z <0.75$ for the benchmark points, including both SM and BSM contributions.} \label{tab:BPratios}
\end{table}

\section{Conclusions}

The existence of portal matter, coupling to both the SM and the dark photon, is a necessary ingredient of the DM kinetic mixing portal scenario. Unlike in the more familiar case where the 
portal matter is a set of vector-like fermions, scalar portal matter fields considered here obtain their physical masses only after the electroweak symmetry breaking associated with the SM and/or $U(1)_D$ 
sectors and so cannot be made arbitrarily heavy. As discussed above, the requirement that after SSB no massless axion-like states be present in the physical spectrum coupled to the 
requirements that the KM mixing parameter(s) be finite and calculable, while the $\rho$ parameter remains essentially unity at tree-level, necessitates the introduction of two new weak 
iso-doublet Higgs representations with opposite dark charges, $Q_D=\pm 1$, as the simplest possibility for the scalar PM model. This implies the new scalar spectrum consists of two 
pairs of charged Higgs states, $H_{1,2}^\pm$, one new CP-odd neutral state $A$, as well as two new CP-even neutral states, $H,h_d$. In this minimal scenario as discussed above, the structure of the extended Higgs potential consistent with all of the gauge symmetries implies that the physical masses of these new, purely electroweak spin-0 states cannot be much 
larger than the SM Higgs vev, \ie,  $\simeq 246$ GeV. In the above scenario, the new scalars we introduce play two essential roles working not only as the PM to generate KM but also 
as the Higgs fields whose vevs are responsible for $U(1)_D$  breaking. 

While the new particles in our model will have rather typical electroweak couplings to the SM $W^\pm, Z$ and 
$\gamma$ gauge bosons, their couplings to the SM fermion fields, as shown above, will be quite highly suppressed by (very) small mixing angle factors. Thus, while these new states 
can be  produced at the LHC in a manner familiar from the examination of the new Higgs fields in many BSM scenarios (which do not involve the SM fermions), their decays will, 
rather uniquely, almost exclusively involve the $W^\pm, Z$ or SM Higgs fields plus either a dark photon or dark Higgs in the final state thus necessarily leading to missing $E_T$ in 
LHC detectors. Amazingly, such new scalar states, though overall rather relatively light on the scale of present day new physics searches, could have up to now evaded the multiple MET 
analyses performed by ATLAS 
and CMS at the LHC in various final state channels for much of the model parameter space as we have demonstrated here. As we seen, in most cases these searches had sensitivities 
rather far from those needed to probe the model space considered here. A few, however, provide the promise of an early window into at least some of this model parameter space once 14 TeV 
LHC running commences. 

One clear way future searches for this model can extend their sensitivities is to lower requirements on MET and/or make greater use of MET significance since, as we have seen, 
the amount of MET is not always large due to both the lightness of these new states and the MET having some tendency to partially cancel in decay/production processes. Although this is 
conventionally somewhat difficult, especially in an even higher luminosity environment with more pileup,  efforts in machine learning may be very helpful here. 

On the theoretical side, one interesting direction for further exploration would be to more completely map out the parameter space allowed by imposing more general copositivity conditions on the Higgs potential of Eq. \ref{higgspot}. In Sec. \ref{sec:unitarity} we restricted ourselves to $c_1=1$ and positive definite $\Lambda$ for simplicity, but a future study may relax these conditions to more fully explore the affine subspace defined by $c_1+c_2=1$ to find matrices which satisfy the more general copositivity conditions outlined in the literature \cite{andersson1995criteria,ping1993criteria}. Another interesting direction for future work is the study of spontaneous CP violation in this model. While above we took the dark vevs $v_{1,2}$ to be real, in general there may be a relative phase between them which may lead to an interesting phenomenology.

\section {Acknowledgements}
This work was supported by the Department of Energy, Contract DE-AC02-76SF00515.

\appendixtitleon
\begin{appendices}
\section{Benchmark Model Points} \label{app:bps}

This Appendix contains the input parameters for each of the four benchmark points described in the text, BP1-BP4. Table \ref{tab:params} lists the values of $t$ and the $\lambda_i$ in the Higgs potential in Eq. \ref{higgspot}. Since the copositivity constraint of Sec \ref{sec:unitarity} forces $\lambda_{67} = \lambda_6 + \textrm{min}(0,\lambda_7)$, we let $\lambda_6=\lambda_{67}$ and take $\lambda_7=1$ for all points. This only impacts $\mathcal{O}(x_i^2)$ terms in the mass relations of Sec. \ref{sec:model} and thus has negligible impact on the analysis of Sec. \ref{sec:signal}.

\begin{table} 
\begin{center}
\begin{tabular}{ |c |c |c |c |c | }
\hline
Parameter & BP1 & BP2 & BP3 & BP4 \\
\hline
$\lambda_1$ & 0.129 & 0.129 & 0.129 & 0.129  \\
$\lambda_{21}$ & 3.8354 & 4.8965 & 4.7086 & 4.9992  \\
$\lambda_{22}$ & 1.5295 & 3.741 & 0.6718 & 1.45263  \\
$\lambda_{31}$ & 4.7436 & 1.5225 & 3.599 & 1.8702 \\
$\lambda_{32}$ & 3.4145 & 1.8583 & 3.2664 & 2.0187  \\
$\lambda_{41}$ & -4.0414 & -1.0256 & -2.5749 & -1.0935  \\
$\lambda_{42}$ & -3.0846 & -1.6024 & -3.2974 & -2.069  \\
$\lambda_{5}$ & -0.5375 & -0.3918 & -0.5768 & -0.3987 \\
$\lambda_{6}$ & 4.095 & 2.7361 & 4.6574 & 3.5631 \\
$\lambda_{7}$ & 1.0 & 1.0 & 1.0 & 1.0  \\
$t = \frac{x_1}{x_2}$ & 1.0773 & 1.1317 & 1.1402 & 1.0838 \\
\hline
\end{tabular}
\end{center}
\caption{Four benchmark points in the parameter space, with the $\lambda_i$ couplings of the Higgs potential in Eq. \ref{higgspot}, and $t$.} \label{tab:params}
\end{table}

The values of the masses and kinetic mixing parameters are listed in Table \ref{tab:BPsum}. As stated in Sec. \ref{sec:scan}, we take $v_1=1$ GeV, and here we use $g_D = e = \sqrt{4 \pi \alpha_{EM}}$ for concreteness in the calculation of $m_V$, $\epsilon$, and $\epsilon_{ZV}$.

\begin{table} 
\begin{center}
\begin{tabular}{ |c |c |c |c |c |c |c |c | }
\hline
Model & $m_H \approx m_A$ & $m_1$ & $m_2$ & $m_{h_d}$ & $m_V$ & $\epsilon$ & $\epsilon_{ZV}$\\
\hline
BP1 & 180.8 GeV & 371.0 GeV & 333.2 GeV & 3.17 GeV & 413.2 MeV & -4.2$\times 10^{-5}$ & -1.9$\times 10^{-4}$ \\
BP2 & 154.7 GeV & 203.9 GeV & 249.0 GeV & 3.37 GeV & 404.1 MeV & 7.7$\times10^{-5}$ & -2.1$\times10^{-4}$ \\
BP3 & 187.8 GeV & 305.6 GeV & 346.2 GeV & 3.27 GeV & 402.8 MeV & 4.8$\times10^{-5}$ & -2.6$\times10^{-4}$  \\
BP4 & 155.7 GeV & 210.5 GeV & 275.3 GeV & 3.28 GeV & 412.0 MeV & 1.0$\times10^{-4}$ & -1.0$\times10^{-4}$  \\
\hline
\end{tabular}
\end{center}
\caption{Four benchmark points in the parameter space, their BSM mass values, and the values of $\epsilon$ and $\epsilon_{ZV}$, assuming $g_D=e$. These points were chosen since they roughly span the range of masses of the scan performed in Sec. \ref{sec:scan}.} \label{tab:BPsum}
\end{table}

\section{$h_{SM} \rightarrow Z V h_d$ Calculation} \label{hzvhdcalc}

Adding the amplitudes of Sec. \ref{sec:raredec}, squaring, and taking the sum over the spin states of the external $Z$, we find

\begin{align*}
\sum_{\textrm{pol.}}\left|\mathcal{M}\right|^2 =& m_Z^2 \tilde \lambda^2 s_{2\theta}^2 \left[ \frac{-c^2 + \frac{(a\cdot c)^2}{m_Z^2}}{[(a+c)^2-m_A^2]^2} + \frac{-b^2 + \frac{(a\cdot b)^2}{m_Z^2}}{[(a+b)^2-m_H^2]^2} + \frac{2(b \cdot c) - 2 \frac{(a \cdot b)(a \cdot c)}{m_Z^2} }{[(a+c)^2-m_A^2][(a+b)^2-m_H^2]} \right]\\
&+\frac{g^2 \tilde \lambda m_Z^2 s_{2\theta} c_{2\theta}}{c_w^2}\left[  \frac{(-c^2+\frac{(a\cdot c)^2}{m_Z^2})(1 - \frac{c^2-b^2}{m_Z^2})+(1+\frac{c^2-b^2}{m_Z^2})(b\cdot c - \frac{(a \cdot b) (a \cdot c)}{m_Z^2})}{[(a+c)^2-m_A^2][(b+c)^2-m_Z^2]} \right] \\
&+\frac{g^2 \tilde \lambda m_Z^2 s_{2\theta} c_{2\theta}}{c_w^2}\left[\frac{(-b^2+\frac{(a\cdot b)^2}{m_Z^2})(1 + \frac{c^2-b^2}{m_Z^2})+(1-\frac{c^2-b^2}{m_Z^2})(b\cdot c - \frac{(a \cdot b) (a \cdot c)}{m_Z^2})}{[(a+b)^2-m_H^2][(b+c)^2-m_Z^2]} \right]\\
+\frac{g^4 c_{2\theta}^2 m_Z^2}{4 c_w^4}&\left[ \frac{(-c^2+\frac{(a\cdot c)^2}{m_Z^2})(1 - \frac{c^2-b^2}{m_Z^2})^2+(-b^2+\frac{(a\cdot b)^2}{m_Z^2})(1 + \frac{c^2-b^2}{m_Z^2})^2+2(1-\frac{[c^2-b^2]^2}{m_Z^4})(b\cdot c - \frac{(a \cdot b) (a \cdot c)}{m_Z^2})}{[(b+c)^2-m_Z^2]^2}   \right],
\end{align*}
where we have defined $a^\mu \equiv p_Z^\mu$, $b^\mu \equiv p_V^\mu$, and $c^\mu \equiv p_{h_d}^\mu$. When we make the approximation of massless $V$ and $h_d$ we then have $b^2=c^2=0$, and $a^2 = m_Z^2$. Integrating over phase space, we can write expressions for the differential width $d\Gamma/d\epsilon_Z$ in terms of $\epsilon_Z \equiv E_z/m_{h_{SM}}$, $x_c = 2 E_{h_d}/m_{h_{SM}}$, and $\mu_Z = m_Z^2/m_{h_{SM}}^2$. Note that we may write $(a\cdot b) = m_{h_{SM}}^2 (1- x_c-\mu_a)/2$, $(a\cdot c) = m_{h_{SM}}^2 (2\epsilon_Z+x_c-1-\mu_a)/2$, and $(b\cdot c) = m_{h_{SM}}^2 (1+\mu_a-2\epsilon_Z)/2$, since $b^2=c^2=0$. We find 
\begin{align*}
\frac{d\Gamma}{d\epsilon_Z} = & \frac{m_{h_{SM}}}{512\pi^3} \int dx_c \bigg\{ \tilde\lambda^2 s2\theta^2 \bigg[ \frac{[2\epsilon_Z+x_c-1-\mu_a]^2}{[2\epsilon_Z+x_c-1-\mu_H]^2} +  \frac{[1-x_c-\mu_a]^2}{[1-x_c-\mu_H]^2}\\ 
& ~~~~~~~~~~~~~~~~~~~~~~~~~~~~~+ \frac{4\mu_a[1+\mu_a-2\epsilon_Z]-2[1-x_c-\mu_a][2\epsilon_z+x_c-1-\mu_a]}{[2\epsilon_Z+x_c-1-\mu_H][1-x_c-\mu_H]} \bigg] \\
&+ \frac{g^2 \tilde \lambda m_Z^2 s_{2\theta} c_{2\theta}}{c_w^2} \bigg[ \frac{[2\epsilon_Z+x_c-1-\mu_Z]^2+2\mu_Z[1-2\epsilon_Z+\mu_Z]-[1-x_c-\mu_Z][2\epsilon_Z+x_c-1-\mu_Z]}{[2\epsilon_Z+x_c-1-\mu_H][1-2\epsilon_Z]} \\
&~~~~~~~~~~~~~~~~~~~~~~~~~~~~+\frac{[1-x_c-\mu_Z]^2+2\mu_Z[1-2\epsilon_Z+\mu_Z] - [1-x_c-\mu_Z][2\epsilon_Z+x_c-1-\mu_Z]}{[1-x_c-\mu_H][1-2\epsilon_Z]} \bigg]\\
+\frac{g^4 c_{2\theta}^2 m_Z^2}{4 c_w^4} & \bigg[ \frac{[2\epsilon_Z+x_c-1-\mu_Z]^2+4\mu_Z[1-2\epsilon_Z+\mu_Z]-2[1-x_c-\mu_Z][2\epsilon_Z+x_c-1-\mu_Z]+[1-x_c-\mu_z]^2}{[1-2\epsilon_Z]^2}  \bigg] \bigg\},
\end{align*}
where $\mu_H \equiv m_H^2/m_{h_{SM}}^2 = m_A^2/m_{h_{SM}}^2$ and $1-\epsilon_Z-\sqrt{\epsilon_Z^2-\mu_Z} \leq x_c \leq 1-\epsilon_Z +\sqrt{\epsilon_Z^2-\mu_Z}$ are the bounds of integration over $x_c$. We have used $2 = 2\epsilon_Z +x_b +x_c$ to replace $x_b$ in the integral. Integrating $d\Gamma/d\epsilon_Z$ with respect to $\epsilon_Z$ over the range $\sqrt{\mu_Z} \leq \epsilon_Z \leq(1+\mu_Z)/2$ then produces the full BSM width.

\end{appendices}

\end{document}